\documentclass[acmsmall]{acmart}

\usepackage{framed}
\usepackage{csquotes}
\usepackage{dirtytalk}
\usepackage{lscape}
\usepackage{comment}
\PassOptionsToPackage{hyphens}{url}\usepackage{hyperref}
\usepackage{appendix}
\usepackage{rotating}
\usepackage{graphicx}
\usepackage{todonotes}
\usepackage{subcaption}
\usepackage{wrapfig}
\usepackage{footmisc}
\usepackage{tabu}
\usepackage{algorithm}
\usepackage{algpseudocode}
\usepackage{multirow}
\usepackage{adjustbox}
\usepackage{amsmath,amsfonts}
\PassOptionsToPackage{normalem}{ulem}
\usepackage{ulem}
\algnewcommand\algorithmicforeach{\textbf{for each}}
\algdef{S}[FOR]{ForEach}[1]{\algorithmicforeach\ #1\ \algorithmicdo}

\definecolor{darkgreen}{RGB}{21,176,26}
\definecolor{orange}{RGB}{242,148,7}

\providecolor{newadded}{rgb}{0,1,0}
\providecolor{omidadded}{rgb}{0,0,1}
\providecolor{deleted}{rgb}{1,0,0}

\graphicspath{{figures/}}

\setcopyright{acmcopyright}
\copyrightyear{2022}
\acmYear{2022}

\acmJournal{TAAS}
\begin{document}

\title{Self-Adaptation in Industry: A Survey}
 
\author{Danny Weyns,$^{1,2}$ Ilias Gerostathopoulos,$^3$ 
Nadeem Abbas,$^2$ 
Jesper Andersson,$^2$ 
Stefan Biffl,$^4$ 
Premek Brada,$^5$ 
Tomas Bures,$^6$ 
Amleto Di Salle,$^7$
Matthias Galster,$^8$
Patricia Lago,$^3$ 
Grace Lewis,$^9$
Marin Litoiu,$^{10}$
Angelika Musil,$^{1,4}$ 
Juergen Musil,$^4$ 
Panos Patros,$^{11}$
Patrizio Pelliccione$^{12}$}
\affiliation{
	\institution{$^1$Katholieke Universiteit Leuven}
	\country{Belgium}
}
\affiliation{
	\institution{$^2$Linnaeus University}
	\country{Sweden}
}
\affiliation{
	\institution{$^3$Vrije Universiteit Amsterdam}
	\country{The Netherlands}
}
\affiliation{
	\institution{$^4$CDL-SQI, TU Wien}
	\country{Austria}
}
\affiliation{
	\institution{$^5$University of West Bohemia}
	\country{Czechia}
}
\affiliation{
	\institution{$^6$Charles University Prague}
	\country{Czechia}
}
\affiliation{
	\institution{$^7$European University of Rome}
	\country{Italy}
}
\affiliation{
	\institution{$^8$University of Canterbury}
	\country{New Zealand}
}
\affiliation{
	\institution{$^9$Carnegie Mellon Software Engineering Institute}
	\country{USA}
}
\affiliation{
	\institution{$^{10}$York University}
	\country{Canada}
}
\affiliation{
	\institution{$^{11}$Raygun Application Performance}
	\country{New Zealand}
}
\affiliation{
	\institution{$^{12}$Gran Sasso Science Institute}
	\country{Italy}
}
\email{danny.weyns@kuleuven.be}

\renewcommand{\shortauthors}{D. Weyns, I. Gerostathopoulos, et al.}

\begin{abstract}
    Computing systems form the backbone  of many areas in our society, from manufacturing to traffic control, healthcare, and financial systems. When software plays a vital role in the design, construction, and operation, these systems are referred as software-intensive systems. Self-adaptation equips a software-intensive system with a feedback loop that either automates tasks that otherwise need to be performed by human operators or deals with uncertain conditions. Such feedback loops have found their way to a variety of practical applications; typical examples are an elastic cloud to adapt computing resources and automated server management to respond quickly to business needs. To gain insight into the motivations for applying self-adaptation in practice, the problems solved using self-adaptation and how these problems are solved, and the difficulties and risks that industry faces in adopting self-adaptation, we performed a large-scale survey. We received 184 valid responses from practitioners spread over 21 countries. Based on the analysis of the survey data, we provide an empirically grounded overview of state-of-the-practice in the application of self-adaptation.  From that, we derive insights for researchers to check their current research with industrial needs, and for practitioners to compare their current practice in applying self-adaptation. These insights also provide opportunities for the application of self-adaptation in practice and pave the way for future industry-research collaborations. 

\end{abstract}




\keywords{self-adaptation, industry, survey}

\maketitle


\section{Introduction}\label{sec:introduction}



Computing systems form the backbone of our factories, traffic control systems, healthcare, telecommunication, financial systems, and so forth. When software plays a vital role in their design, 
 construction, and operation, these systems are often referred to as software-intensive systems~\cite{Holzl2008}. The trustworthiness and sustainability of these systems is vital for our society~\cite{2889160.2889212,978-3-319-48992-6_1}. Yet, building and maintaining trustworthy and sustainable systems is challenging due to complexity that arises from the growing demands on these systems, their continued integration, the uncertain operating conditions they face, the fast speed of technological progress, etc. These challenges have been a continuous driver for new and innovative approaches to design, develop, and operate software-intensive systems. One common approach today is so called DevOps in which development and operation are blended, allowing system components to be easily evolved and redeployed without impacting their operation~\cite{cheng2009software}. 

A classic approach to address the increasing complexity of software-intensive systems is transferring control from humans~\cite{Lethbridge2005SSE} to software components by equipping systems with feedback loops that automate tasks that otherwise need to be performed by human operators. 
These feedback loops monitor the system and its environment, reason about the system behaviour and its goals, and adapt the system to ensure its goals under changing conditions, or gracefully degrade if necessary. Such goals can be very diverse, ranging from ensuring a required level of performance under uncertain workload conditions, dealing with errors caused by external services that are difficult to predict, or defending the system against malicious attacks and the problems they may cause. A typical example is a feedback loop deployed in a cloud environment that expands or decreases computing resources to meet changing demands while minimising the cost of operation. Another example is a container framework that performs autoscaling in a microservice deployment. 

The principles of applying feedback control to software-intensive systems have been the subject of active study in academia. Back in 1998, Oreizy et. al.~\cite{oreizy1999aba} presented a seminal paper at the International Conference on Software Engineering (ICSE) where the authors introduced the notion of \textit{self-adaptation} that comprises two simultaneous processes: system adaptation that is concerned with detecting and handling changing circumstances, and system evolution that is concerned with the consistent application of change over time. A few years later, Garlan et. al.~\cite{garlan2004rainbow} stated the crucial role of architectural models as first-class citizens that enable a system to reason about system-wide change and adapt itself accordingly to achieve or maintain its goals. Blair et. al.~\cite{Blair2009MR} consolidated and elaborated on these principles in what is now generally known as ``models at runtime.'' In 2007, Kramer and Magee~\cite{Kramer2007SMS} stated the crucial role of software architecture in the realisation of self-adaptive systems, distinguishing adaptation management from goal management. Over the last decade, the research community has developed a vast body of knowledge and know-how on principles, see e.g.,~\cite{andersson2009modeling,Blair2009MR,FilieriICSE14,978-3-319-74183-3-12}, models and languages~\cite{Relax2010Whittle,2593929.2593944,2555612,MetzgerQMBP20}, processes and methods~\cite{Andersson2013,Cheng2014,8008800}, patterns~\cite{1808984.1808990,weyns2013patterns,9223653}, and frameworks~\cite{garlan2004rainbow,Rouvoy2009,1882291.1882296} to engineer self-adaptive systems. Researchers have documented a substantial number of literature reviews and surveys on various topics in self-adaptive systems, such as the benefits of self-adaptation~\cite{6224395}, requirements for self-adaptive systems~\cite{978-3-319-05843-65},  approaches to realise self-adaptation~\cite{MACIASESCRIVA20137267,MAHDAVIHEZAVEHI20171,9223653,7929422}, the use of formal methods in self-adaptive systems~\cite{2347583.2347592}, self-protection~\cite{2555611},  
the notion of uncertainty~\cite{MAHDAVIHEZAVEHI201745,10.1145-3487921}, and the use of machine learning in the realisation of self-adaptation~\cite{3469440}, among others. Basic research works in the field of self-adaptation are for example~\cite{huebscher2008survey,1516533.1516538,cheng2009software,Lemos2013roadmap,weyns2021introduction}. 

In parallel, the principles of feedback control have been studied and applied in  industry. For example, about two decades ago, IBM launched its legendary initiative on  autonomic computing~\cite{kephart2003vision}. Inspired by the autonomic nervous system of the human body, the central idea of autonomic computing was to enable computing systems to manage themselves based on high-level goals. Four classic goals are self-optimisation, self-healing, self-protection, and self-configuration. Autonomic computing delegates the complexity of system operation to the machine aiming to reduce the time required by operators to resolve system difficulties and other maintenance tasks such as software updates. Over the years, industrial solutions based on feedback loops have found their way to practical applications, for instance in the domain of elastic cloud to adapt computing resources and automated management of server parks to deal with changing business needs,  e.g.,\,\cite{beyer2016,spyker2020}.

While the output of academic research is documented in research articles, journal volumes, and books, the current practice of self-adaptation in industry has never been systematically described. 

\subsection{Objective and Research Questions}
\label{subsec:research_objective_and_rqs}

Our general objective is to better understand the state of practice of self-adaptation in industry. To that end, we perform a large-scale survey with active practitioners. Concretely, this survey aims at shining a light on what motivates practitioners to apply self-adaptation, what kind of problems they solve using self-adaptation, how practitioners design and develop self-adaptive systems, whether they follow any established practices, what difficulties and risks they face in adopting self-adaptation, and what future opportunities industry sees for the application of self-adaptation. 

To the best of our knowledge, no systematic study has been done that investigates and these issues. Hence, there is no clear and documented view of why and how the principles of self-adaptation are applied in practice, and what challenges practitioners face when realising self-adaptation.
Investigating industrial practice on self-adaptation and answering the questions targeted by this study will help narrow the gap between industry and academia. It aims at helping researchers in academia to get a better picture of how self-adaptation is applied in practice, the industrial needs in realising self-adaptation, and what problems practitioners face. We conjecture that having a better picture about industry practice will help the research community to position their efforts with respect to industrial needs and make well-informed decisions to set future research objectives, both fundamental and applied. On the other hand, drawing a picture of the state-of-the-practice can also benefit industry by sharing the motivations and potential benefits of self-adaptation, directing them towards relevant sources of information such as best practices, 
and identifying opportunities for collaboration with researchers to address the problems they face. 

We aim to answer the following concrete research questions:

\begin{description}
    \item[RQ1:] What drives practitioners to apply self-adaptation in software-intensive systems?   
    \item[RQ2:] How do practitioners characterise self-adaptation?
    \item[RQ3:] How do practitioners apply self-adaptation in industrial software-intensive systems?
    \item[RQ4:] What are the experiences of practitioners with applying self-adaptation and do they see opportunities for how and where to apply self-adaptation? 
\end{description}

With RQ1, we want to investigate the motivations of practitioners for applying self-adaptation, the kinds of  industrial systems for which self-adaptation is applied, and the types of problems they solve using self-adaptation. 
In academic research, self-adaptation has been proposed for two main complementary problems\,\cite{weyns2021introduction}: 1) to automate the management of complex software-intensive systems based on high-level goals provided by operators, and 2) to deal with operating conditions that are hard to predict before deployment and need to be resolved during operation (i.e., mitigating uncertainties). Key management tasks for self-adaptation are self-healing, self-optimisation, self-protection, and self-configuration. We want to understand whether industry uses the principles of self-adaptation to deal with the same or different problems, and whether and how they relate to the classic system and software management tasks. Answering RQ1 will shine a light on application areas, motivations, and concrete problems for which self-adaptation is applied by practitioners or could be applied by practitioners who currently do not use self-adaptation. This may provide academics with insights in relevant areas to drive and validate research results on self-adaptation. The results may also indicate applications and problems that are not yet explored in industry and may benefit both academia and industry.  

With RQ2, we aim to investigate the perception of practitioners on the concept of \emph{self-adaptation}. We are particularly interested in how practitioners characterise self-adaptation as a property that enables a system to adapt itself at runtime. 
To that end, we will elicit concrete examples of what they understand by self-adaptation. This will give us better understanding of whether and how practitioners understand the concept of self-adaptation, what terminology they use, whether there are any differences in the viewpoints on what constitutes self-adaptation, and whether they consider self-adaptation altogether useful. This may also shine a light on whether there are any (emerging) industrial standard practices, e.g., a \,technology stack or tools.
Answering RQ2 will help researchers to get a better picture of how practitioners understand the concept of self-adaptation. On the other hand, the insights may reveal potential opportunities for practitioners to benefit from expertise of other practitioners as well as knowledge developed by researchers. 

With RQ3, we aim at examining how self-adaptation has been realised and used in industry. We are particularly interested in mechanisms, tools, benchmarks, and processes employed in the industry to engineer self-adaptive solutions. 
We will pay attention to the degree of automation and the role of humans in runtime adaptation as this is commonly considered important for trust in software-intensive systems, see e.g.,\,\cite{978-3-030-00761-4-4}. 
Furthermore, we are interested in comparing industrial practices with solutions developed by academics, such as modelling techniques, frameworks, and verification techniques. We also want to understand how practitioners  obtain  trust  in  the  self-adaptive  solutions  they employ.  Answering RQ3 will provide insights into best practices on how practitioners realise self-adaptation. It will highlight the criteria that practitioners use to apply and realise self-adaptation solutions and may shine a light on to what extent solutions from the research community have been adopted in  industry. These insights will open opportunities for both academia and industry to steer future research and improve practical applications.  

Finally, with RQ4, we want to understand the difficulties and risks, if any, that practitioners experience in the design, implementation, and other engineering activities of self-adaptive systems. We also will probe whether practitioners face problems for which they would appreciate support from researchers. Finally, we elicit opportunities that practitioners see for applying self-adaptation  that are not exploited yet. Answering RQ4 may help to fill the gap between academia and industry. Furthermore, identifying problems and risks may trigger new collaborative studies to investigate and address these challenges. Such studies are likely to bridge the gap and result in more targeted research and improved industrial applications of self-adaptive systems.

\subsection{Contributions}
\label{subsec:contributions}
By drawing a landscape of the use of self-adaptation in industry, the survey results benefit both researchers and practitioners. Concretely, the contributions of this study are: 
\begin{itemize}
    \item An empirically grounded overview of state-of-the-practice in the application of self-adaptation; 
    \item Insights for researchers to assess their current research in relation to industrial needs;
    \item Insights for practitioners to assess the level of their current practice in applying self-adaptation; 
    \item Additional prospects for applying self-adaptation in practice and opportunities for industry-research collaborations. 
\end{itemize}

Preliminary results of this study were reported in\,\cite{3524844.3528077}. That paper only considered a small subset of questions and reported initial results based on one batch of data. 


\subsection{Outline}
\label{sec:paper_outline}
In Section~\ref{sec:research_method} we present the study design with the  survey questions and analysis methods used. Section~\ref{sec:results} presents the results for each research question and provides key insights for each research question. In Section~\ref{sec:cross_analysis}, we derive insights from the study results for researchers and practitioners. Section~\ref{sec:ThreatsToValidity} discusses threats to validity. Finally, we wrap up and conclude in Section~\ref{sec:conclusions}.

\section{Research Method}
\label{sec:research_method}

In this study we use a survey as research method~\cite{gray2013drr}. Subsequently, we discuss the population and sample, the questionnaire, and the data analysis methods we used.

\subsection{Population and Sampling}\label{subsec:population_sampling}

Our target population are practitioners that are actively involved in the engineering of industrial software-intensive systems in any domain. This includes architects, designers, developers, testers, maintainers, operators, and other people who have technical expertise and are actively involved in the development and maintenance of these software systems. 

Concretely, we contacted 355 practitioners from a wide variety of companies\footnote{Almost all practitioners we contacted were from different companies and the few that were from the same company had different roles within the company. The participants were asked to answer from their own perspective.} via the networks of the researchers involved in this study (i.e., the authors of this paper) to complete the survey. We used two criteria to invite people: (1) participants should be active in different domains that are representative of software-intensive systems, and (2) participants have the required expertise to answer the questions. The invited practitioners were spread over in total 21 countries.\footnote{Sweden 58 invited practitioners, USA 55, Austria 50, Belgium 42, Czech Republic 34, Germany 23, New Zealand 22, The Netherlands 18, Canada 15, Spain 9, Denmark 7, UK 5, France 4, India 2, Greece 2, Poland 1, Norway 1, Switzerland 1, Australia 1, Japan 1, Unknown 2} The invitations were sent by personalised emails in different batches during the period from November 30, 2020 until July 31, 2022. We sent reminders according to a predefined \mbox{schedule of one, two, and six weeks after the invitation.} 

\subsection{Survey Instrument}

The survey used a questionnaire to collect data based on a set of predefined questions~\cite{gray2013drr}. 
Because practitioners are not necessarily familiar with the term self-adaptation, the survey started with a gentle introduction of the core idea of what constitutes a self-adaptive system using basic terminology commonly used in industry, and illustrated this with a few characteristic examples to make it concrete. 
We used both closed and open questions. Closed questions have a predefined set of answers, such as yes/no, or multiple choice. We also allowed participants to add extra options for answering several closed questions using a text field. Open questions provide a space that participants can use to provide an answer. While closed questions allow acquiring a clear view on a particular topic using basic statistics, open questions allow acquiring in-depth insights using qualitative analysis. 
We provide a replication package with all study materials, including the study protocol, the questionnaire, the raw data, and the analysis results.\footnote{https://people.cs.kuleuven.be/danny.weyns/surveys/sas-in-industry/}

For this study we used a self-administered anonymous online questionnaire (Survey \& Report hosted by Linnaeus University, Sweden). 
The main motivation to use an online questionnaire is to involve a large set of participants with relatively low cost (both time-wise and financially). 
We created an initial list of survey questions that were directly derived from the research questions of this study. The initial list of questions was composed by two members of the research team and then crosschecked by the other team members. 

We validated the questionnaire in a pilot with eight randomly-selected participants from the target population. For this pilot, we added additional meta-questions to the questionnaire about clarity of terminology and questions, relevance of the questions, scope of the questions, and the time required to complete the survey. For both clarity of terminology and clarity of the questions we obtained an average score of 4.38 on a scale from 1 (Not clear at all) to 5 (Very clear). None of the participants indicated that questions should be removed or modified. Six participants indicated that no important aspects were missing. One participant hinted that we may also probe whether the use of self-adaptation requires a specialised team in the company or alternatively infrastructure to share knowledge. Another participant suggested adding a question about scalability of solutions for self-adaptation. One participant stated that the example system we used to introduce self-adaptation may create some bias, and further that answers to questions may differ depending on roles on the engineering teams. The average reported time to complete the survey was 24 minutes. Based on the feedback, we adjusted the introductory part of the questionnaire. We did not revise the questions as they were perceived as clear and well scoped. The finalised questionnaire was then distributed to the participants as explained above. 

The first part of the questionnaire (Table~\ref{tab:demographics}) solicited whether the participant applies self-adaptation and collected general demographic information. This allowed us to check whether the participant had experience with self-adaptation (Q0.1), confirm a good coverage of kinds of software-intensive systems across participants (Q0.2), the size of the companies of participants (Q0.3), as well as a confirmation of the participant's role (Q0.4) and years of experience (Q0.5).  

\small
\begin{table}[!h]
\caption{Questionnaire: Demographic information}
\label{tab:demographics}
\begin{tabular}{lp{6.6cm}p{5.4cm}}
\hline\noalign{\smallskip}
ID & Question & Response options \\
\noalign{\smallskip}\hline\noalign{\smallskip}
Q0.1 & Have  you  worked  with  concrete  self-adaptive  systems? & Yes; No \\
Q0.2 & What kind of software systems does your organisation build? & Free text \\
Q0.3 & Approximately, how many people are working on engineering software in your organisation? & 1-10; 11-20; 21-50; 51-100; more than 100\\
Q0.4 & What is your role in your organisation? & Project Manager; Designer; Programmer; Tester; Operator; Maintainer; Other (free text)\\
Q0.5 & How many years of software engineering experience do you have in total? & 1-3 Years; 4-8 Years; 9-20 Years; If other, please specify (free text) \\
\noalign{\smallskip}\hline
\end{tabular}
\end{table}
\normalsize

The second part of the questionnaire aimed at questions related to RQ1 collecting data about the problems for which the participants apply self-adaptation (Q1.1), the main business motivations for using self-adaptation (Q1.2), and the benefits obtained from applying self-adaptation (Q1.3) (see Table~\ref{tab:rq1}). The first two questions had multiple options.\footnote{Question Q1.2 aimed at investigating motives for applying self-adaptation at a more high-level, whereas Q1.3 was focusing more at low-level benefits, technical and specific to a system. The initial lists of the options for these questions were based on the literature of self-adaptation, see e.g.,~\cite{Lemos2013roadmap,Weyns2013,9462043}.}  

\small
\begin{table}[hbt]
\caption{Questionnaire: Drivers for applying self-adaptation in industrial software-intensive systems (RQ1)}
\label{tab:rq1}
\begin{tabular}{lp{6.6cm}p{5.4cm}}
\hline\noalign{\smallskip}
ID & Question & Response options\\
\noalign{\smallskip}\hline\noalign{\smallskip}
Q1.1 & For which problems do you or your organisation apply self-adaptation capabilities, i.e., a managing system that monitors and adapts a managed system to achieve some objectives? & To automate tasks; To deal with changes in the environment; To deal with changes in business goals; To optimise system performance; To detect and resolve errors; To detect and protect a system against threats; To configure/reconfigure a system; Other (free text)\\
Q1.2 & What are the main business motivations for you or your organisation to apply self-adaptation? & To improve user satisfaction; To reduce costs; To mitigate risks; To open up new application opportunities; Other (free text)\\
Q1.3 & What could be the benefit of self-adaptation in one of the systems you worked with? Please explain briefly. & Free text\\


 \noalign{\smallskip}\hline
\end{tabular}
\end{table}
\normalsize

The third part of the questionnaire covered a question related to RQ2 on how practitioners characterise self-adaptation (see Table~\ref{tab:rq2}). This part included only one question that asked participants to describe a concrete self-adaptive system they had  worked with (Q2.1).  

\small
\begin{table}[hbt]
\caption{Questionnaire: How practitioners characterise self-adaptation (RQ2)}
\label{tab:rq2}
\begin{tabular}{lp{8.6cm}p{3.4cm}}
\hline\noalign{\smallskip}
ID & Question & Response options\\
\noalign{\smallskip}\hline\noalign{\smallskip}
Q2.1 & Think of a concrete self-adaptive system you worked with. Name this system and briefly explain its purpose (please use this system to answer the following three questions) & Free text\\

\noalign{\smallskip}\hline
\end{tabular}
\end{table}

\small
\begin{table}[h!]
\caption{Questionnaire: How self-adaptation is applied in industrial software-intensive systems (RQ3)}
\label{tab:rq3}
\begin{tabular}{lp{8.6cm}p{3.4cm}}
\hline\noalign{\smallskip}
ID & Question & Response options\\
\noalign{\smallskip}\hline\noalign{\smallskip}
Q3.1 & What mechanisms or tools does the self-adaptive system you worked with use to monitor a managed system during operation? By monitor, we mean tracking properties of the system or its environment. & Free text\\
Q3.2 & What mechanisms or tools does the self-adaptive system you worked with use to analyse conditions of a managed system during operation? By analyse, we mean examining conditions of the system or its environment and determining whether any adaptation is required or not. & Free text\\
Q3.3 & What mechanisms or tools does the self-adaptive system you worked with use to change a managed system or parts of it during operation? By change, we mean adjusting parameters of the system, or adding, removing or changing any parts of it. & Free text\\
Q3.4 & What is the degree of automation of the majority of the self-adaptive solutions you work with in your organisation? & Semi-automated; Fully automated; Mixed (Semi and Fully Automated); Other (free text)\\
Q3.5 & Do you reuse solutions to realise self-adaptation in systems you work with? & Never; Very Rarely; Rarely; Sometimes; Frequently; Very Frequently; Always \\
Q3.6 & Please provide a concrete example of reuse you used to realise self-adaptation? & Free text \\
Q3.7 & Why do you not often reuse solutions when realising self-adaptive systems? What hinders the reuse, please provide a short answer. & Free text\\
Q3.8 & How do you ensure that you can trust the self-adaptive solutions you build? Examples could be extensive testing or human supervision, but you may use other means. Please describe briefly. & Free text\\
\noalign{\smallskip}\hline
\end{tabular}
\end{table}
\normalsize

The fourth part of the questionnaire addressed RQ3 on how practitioners apply self-adaptation in their practice (see Table~\ref{tab:rq3}). The first three questions investigated the mechanisms that participants use to monitor (Q3.1) and analyse (Q3.2) the system during operation, and change the system when needed (Q3.3). The next question investigated the degree   of automation of self-adaptation (Q3.4). The next three questions investigate reuse of solutions (Q3.5-Q3.7). The last question of this part of the questionnaire probed whether and how practitioners establish trust in the self-adaptation solutions they build (Q3.8).  

\small
\begin{table}[h!]
\caption{Questionnaire: Risks, challenges, and opportunities when applying self-adaptation in practice (RQ4)}
\label{tab:rq4}
\begin{tabular}{lp{8.6cm}p{3.4cm}}
\hline\noalign{\smallskip}
ID & Question & Response options\\
\noalign{\smallskip}\hline\noalign{\smallskip}
Q4.1 & Did you encounter particular difficulties or challenges when engineering or maintaining self-adaptive systems you worked with? & Never; Very Rarely; Rarely; Sometimes; Frequently; Very Frequently; Always\\
Q4.2 & Please give one or two examples of the difficulties or challenges that you encountered when engineering or maintaining self-adaptive systems.  & Free text\\
Q4.3 & Did you face any risks when engineering self-adaptive systems you worked with? & Never; Very Rarely; Rarely; Sometimes; Frequently; Very Frequently; Always\\
Q4.4 & Please briefly describe one or two risks that you faced when engineering self-adaptive systems. & Free text\\
Q4.5 & How did you mitigate the risks that you faced? Please explain briefly. & Free text\\
Q4.6 & Have you faced or seen any problems of self-adaptation for which you would appreciate support from researchers? & Never; Very Rarely; Rarely; Sometimes; Frequently; Very Frequently; Always\\
Q4.7 & For which problems of self-adaptation would you appreciate support from researchers? Please briefly explain one or two such problems. & Free text\\
Q4.8 & In your organisation or in industry in general, do you see application opportunities for self-adaptation that are currently not exploited? & Yes; No\\
Q4.9 & Please describe or give examples of the application opportunities for self-adaptation that are currently not exploited. & Free text\\
\noalign{\smallskip}\hline
\end{tabular}
\end{table}
\normalsize

Finally, the fifth part of the questionnaire addressed RQ4 on difficulties, risks, and opportunities of applying self-adaptation in practice (see Table~\ref{tab:rq4}). The first two questions investigated difficulties (Q4.1 and Q4.2); the next three questions focused on risks and risk mitigation (Q4.3-Q4.5). The next two questions probed the interest of practitioners to get support from researchers for solving problems with self-adaptation (Q4.6 and Q4.7). The last two questions investigated opportunities for applying self-adaptation beyond the current practice (Q4.8 and Q4.9).  



The questionnaire concluded with a question (Q5.1) about how confident participants were in general about the answers they gave when answering the survey questions with possible answers: Very confident; Confident; Sufficiently confident; Neutral; Somewhat unconfident; Not confident; Not confident at all.

\subsection{Data Analysis}

To analyse closed questions, we used descriptive statistics and quantitative data analysis. Therefore, we mostly report frequencies of answers,  percentages relative to the respective number of responses, and relationships between answers to questions based on contingency matrices (based on the categorisation of answers). We only report relationships that led to relevant insights. 

To analyse comments to open questions, we used qualitative data analysis. In particular, we used inductive reasoning to construct codes and  infer categories from the data by labelling occurrences of codes and grouping them into categories~\cite{Stol2016}. Similar to others (e.g., Prechelt et al.~\cite{Prechelt2018}), we tried to keep coding simple. We did not have a pre-defined coding schema or a pre-defined granularity or semantic style for the codes. However, we interpreted comments in the context of the question for which they were given.  We used a simple version of open coding~\cite{Strauss1990}. Similar to Mendez Fernandez et al.~\cite{Fernandez2016}, we used open coding to add codes to small coherent fragments of the comments. We then categorised the developed concepts in a hierarchy of categories as an abstraction of the codes. We coded in sub-teams of two or three coders in total 886 comments of 12 open questions. Coding was first done individually and then consolidated in the sub-team. Two other researchers crosschecked the consolidated coding. Where necessary, the coding was adjusted in consensus between the sub-team and the researchers. We excluded some comments from coding, e.g., if they did not provide any additional insights or if they were too generic, e.g., a participant answering ``Always'' to a closed question and stating ``This is how we work'' in the comments. Also, we did not map the answers to a closed question to comments for that question. For example, a participant may have answered that they never reused solutions for self-adaptation,  but in their comments indicated reasons that they ``might'' do so (i.e., one comment may cover several concepts, which may not necessarily match the answer to the closed question). When reporting example quotes from comments in Section~\ref{sec:results}, we use verbatim excerpts, including spelling and punctuation errors.

\section{Results}
\label{sec:results}

\subsection{Demographic Information}
\label{subsec:demographic_information}
In total, 184 participants completed the survey from 355 invitations, i.e., a response rate of 51.8\%. 
Based on the answers to the first question (Q0.1), we split the answers of the other questions of the demographics in two groups: those provided by all participants and those provided by participants that worked with concrete self-adaptive systems.\footnote{While we selected participants that have the required expertise to answer questions (second criterion in Section\,\ref{subsec:population_sampling}),  this does not necessarily mean that they have worked (or are working) with concrete self-adaptive systems.}

\subsubsection{Experience with self-adaptation (Q0.1):} Of the 184 participants that provided valid data, 100 (54.4\%) expressed to have worked with concrete self-adaptive systems.
\subsubsection{Software systems built by organisations (Q0.2):} Almost all participants (181, 98.4\%) provided a valid description of the kind of systems they build. 
Based on the analysis of the data we could classify the answers along two axes: the \textit{types} of software systems built by the organisations, and the \textit{focus} of the software systems. The type refers to the domain, while the focus refers to the activities on which the organisation concentrates within the domain. For example, automation (focus) within manufacturing (type). Note that the domain may be abstract, e.g., embedded systems or communication and networks. Focus on the other hand may refer to purpose, such as analytics, but also specific technologies or methods, such as machine learning. 

\begin{figure}[h]
\centering
\includegraphics[width=\columnwidth]{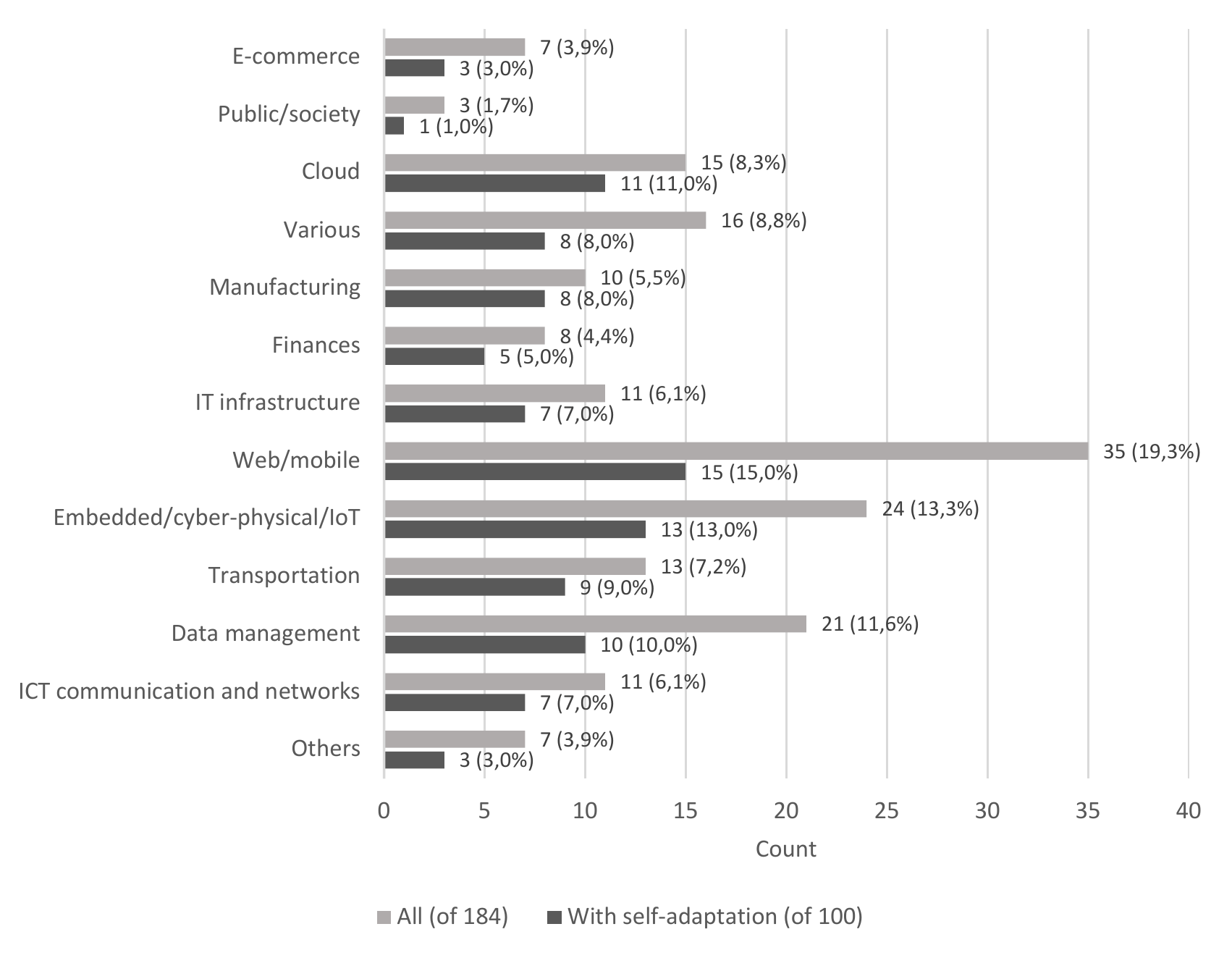}\vspace{-10pt}
\caption{Types of software systems build by organisations (Q0.2).}\vspace{-10pt}
\label{tab:q1-1-types}
\end{figure}

Figure~\ref{tab:q1-1-types}\footnote{Because the number of participants that worked with self-adaptive systems is 100 and all provided a valid description, the absolute numbers are also percentages. We also apply this to the data of the other questions unless differently stated.} summarises the types of systems we identified. The most frequent types are web/mobile, embedded, cyber-physical, IoT systems, data management, and cloud (together these four types represent 52.5\% of all systems). Sixteen participants (8.8\%) built various types of systems.\footnote{The option ''Various'' refers to different kind of systems. The option ''Others'' on the other hand refers to specific types of systems different from those listed in the table, e.g., a system for grading software at educational institutions.} The results show that the percentages of the types of systems of all participants and those that worked with self-adaptation are similar. 

\begin{figure}[h]
\centering
\includegraphics[width=0.9\columnwidth]{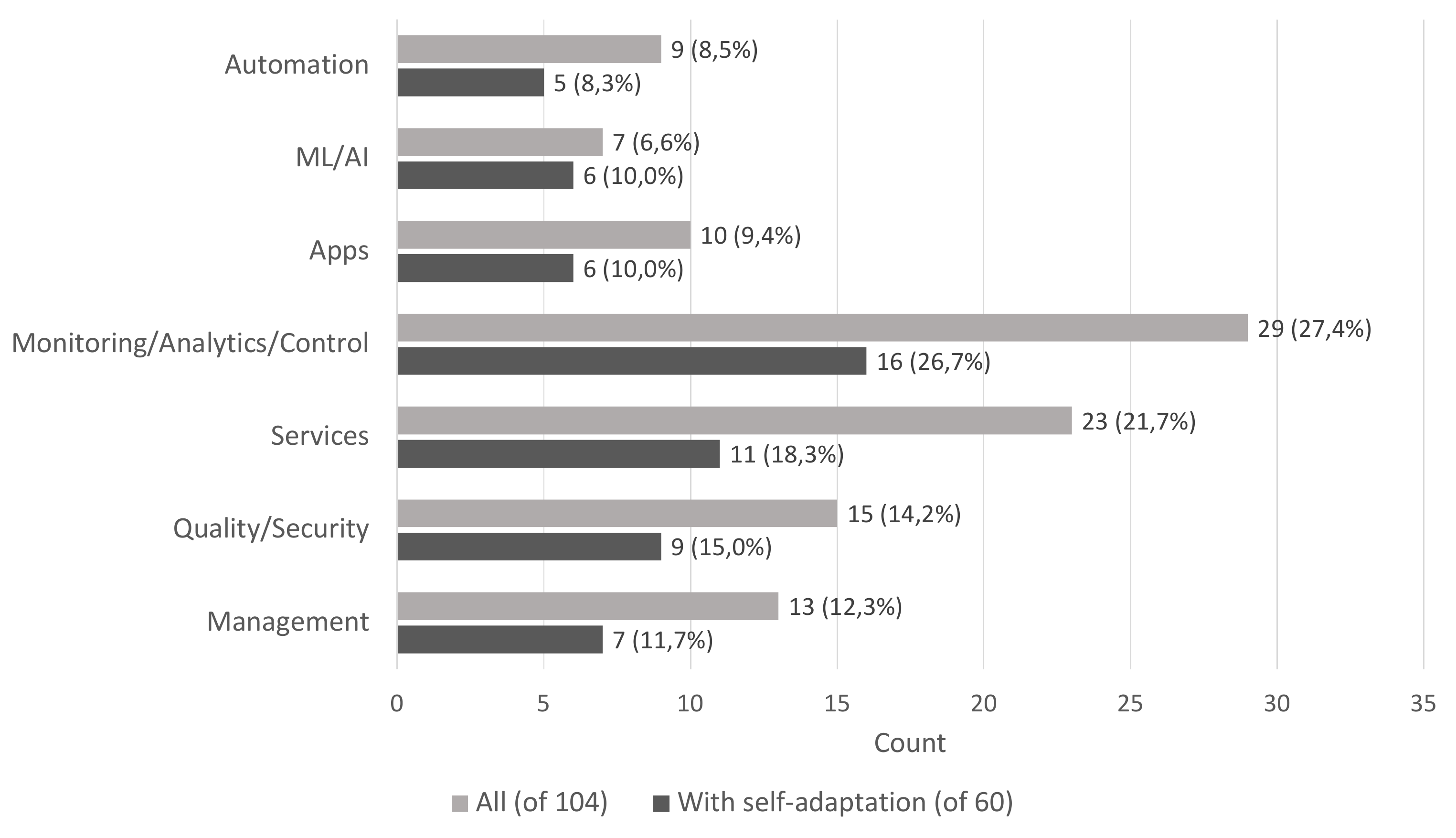}
\vspace{-10pt}
\caption{Focus of software systems built by organisations (Q0.2).}\vspace{-10pt}
\label{tab:q1-1-focus}
\end{figure}

In addition to the types of systems, 104 participants (56.5\%) also provided insights in the focus of the systems they build. Among the 100 participants that worked with self-adaptation, 60 provided a description of the focus. 
Figure~\ref{tab:q1-1-focus} 
provides an overview the results. The dominant focus is monitoring, analytics and control, representing 27.4\% of the foci described by the participants. Other key foci are services (21.7\% of the participants that described the focus of the systems they built) and quality and security (14.2\%).  
Overall, the variety in the types of systems built by the participants and the different foci in activities underpins the representativeness of the data collected during the survey.  

\subsubsection{Software engineers working at companies (Q0.3):} 
\begin{figure}[h]
\centering
\includegraphics[width=0.7\columnwidth]{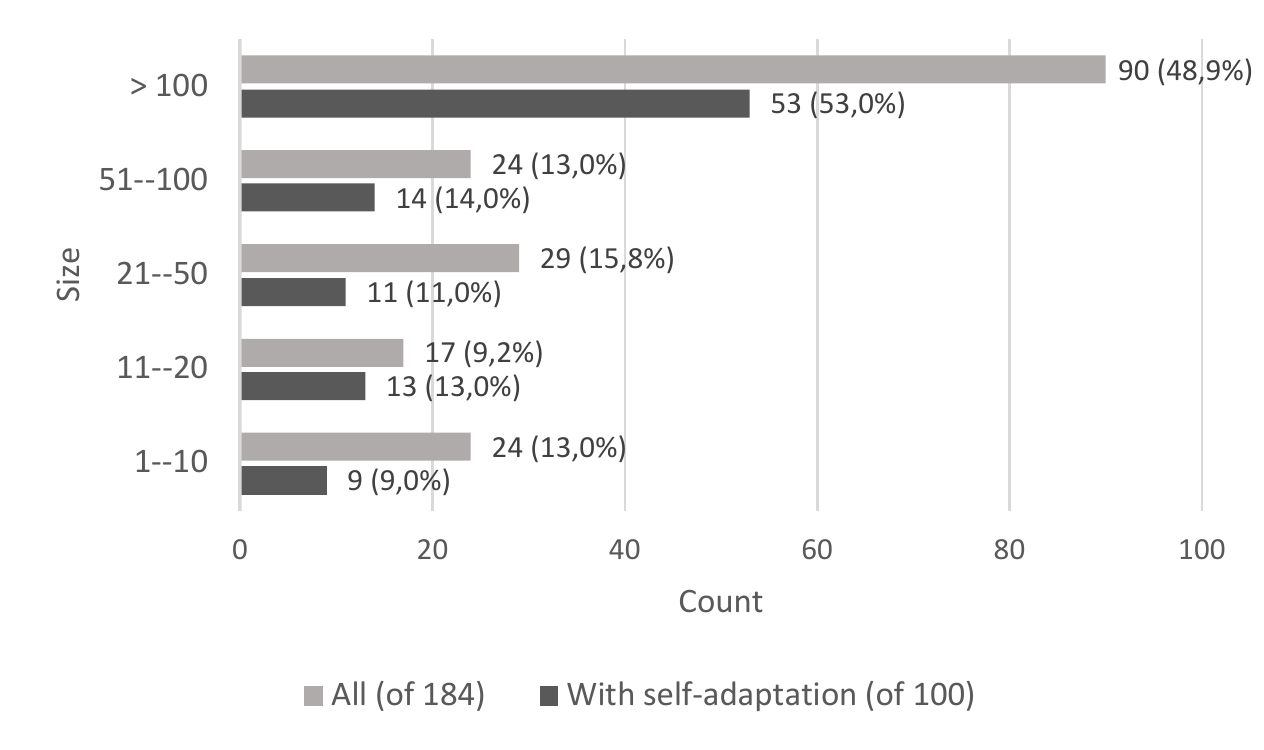}\vspace{-10pt}
\caption{Size of companies (Q0.3).}\vspace{-10pt}
\label{fig:q0-3}
\end{figure}

Figure~\ref{fig:q0-3} summarises the results of the number of software engineers that work at the companies of the  participants. About half of the companies have more than 100 employees who work as software engineers. The other half is about equally divided over four categories of companies with between 1 and 100 software engineers. The results are similar for all participants and those that have worked with  self-adaptive systems. The numbers show that we collected data from participants of companies with a significant number of software engineers, i.e., people dedicated to building software-intensive systems (because our study is interested in the engineering of software-intensive systems, we collected the number of software engineers at the companies and not the total number of employees as a measure for size). 

\subsubsection{Roles of participants in their organisation (Q0.4):} 
\begin{figure}[h]
\centering
\includegraphics[width=0.9\columnwidth]{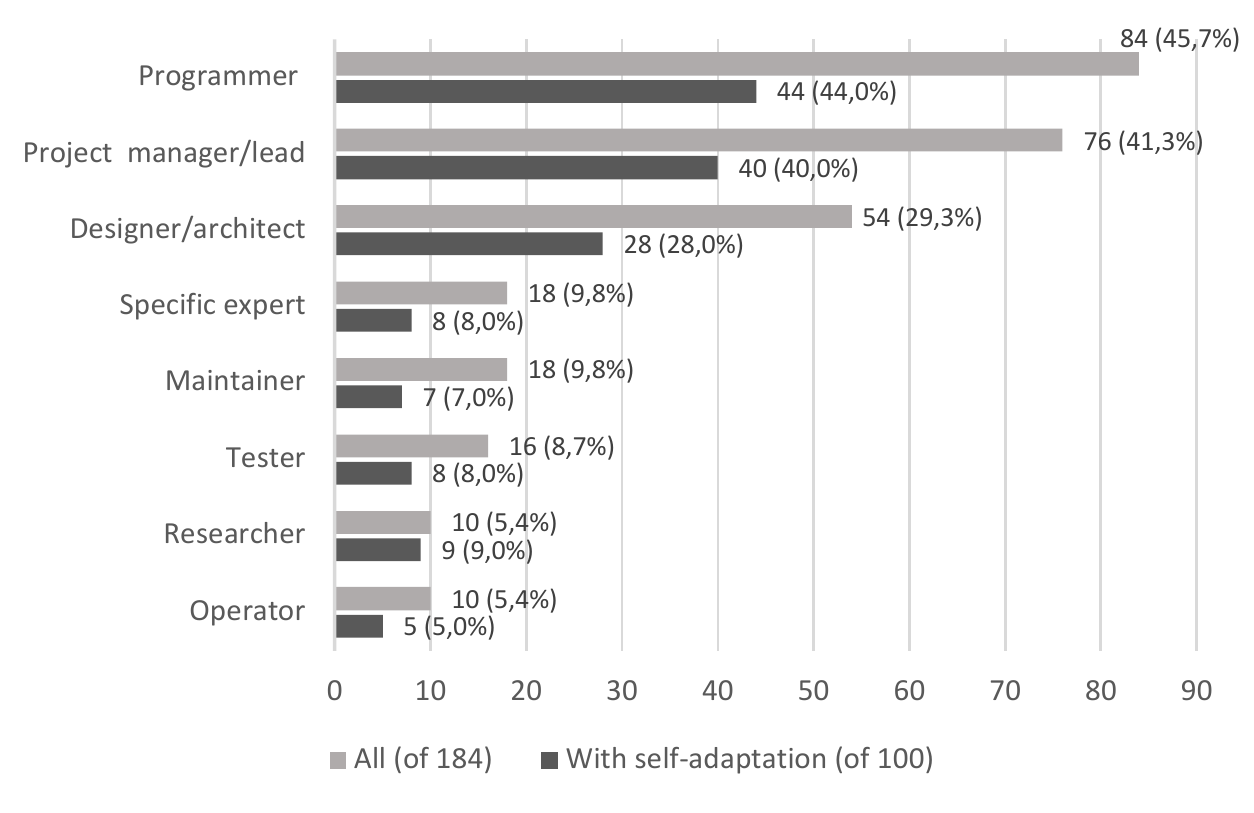}\vspace{-10pt}
\caption{Roles of participants in their companies (Q0.4).}
\label{fig:q0-4}
\end{figure}

The role(s) that participants have in their company are summarised in Figure~\ref{fig:q0-4}. Of 184 participants, 129 indicated that they have one role in their organisation. The other participants indicated that they have two or more roles. 
Overall, the participants reported on average 1.6 roles in their company. The participants that worked with self-adaptation reported on average 1.5 roles.  The most frequent roles are programmer and project manager/coordinator, each representing over 40\% of the participants. About one in three participants works as a designer or architect. The representation of the other roles is lower in the sample. The relative numbers for the roles of all participants and those that work with self-adaptive systems are again similar. One exception is researcher: 9 of the 10 participants that work as researcher have worked with self-adaptive systems. The results show that we collected data from participants with a broad range of key software engineering roles in industry. 

\subsubsection{Experience of participants (Q0.5):} 
\begin{figure}[h]
\centering
\includegraphics[width=0.7\columnwidth]{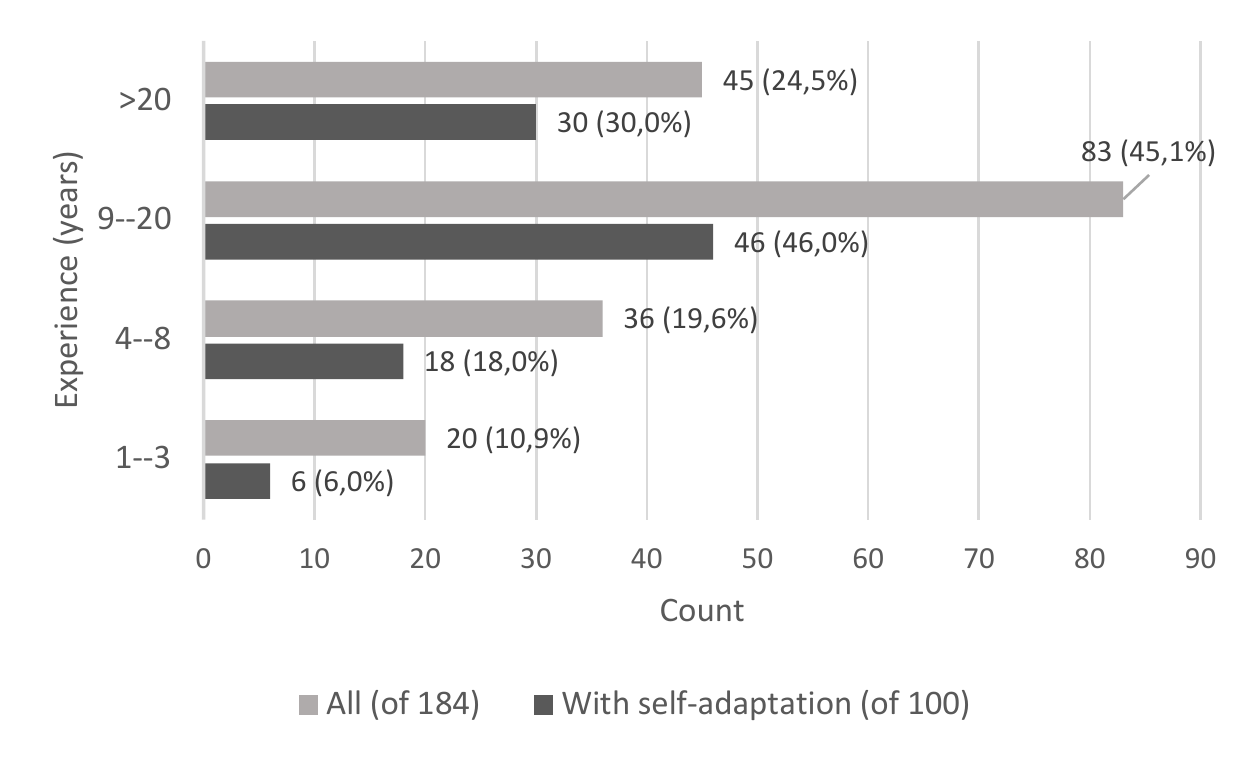}\vspace{-10pt}
\caption{Software Engineering experience of participants (Q0.5).}
\label{fig:q0-5}
\end{figure}

Figure~\ref{fig:q0-5} summarises the years of experience of participants as software engineers.\footnote{Expertise can be based on any role in relation to engineering software-intensive systems as shown in Table~\ref{tab:q0-3}.} A majority of the participants have at least 9 years of experience as software engineer; i.e., 69.6\% of the total sample and 76.0\% of the practitioners that worked with self-adaptation. The distributions for all the participants and those that worked with self-adaptation are similar. The numbers show that most participants of the survey are experienced software engineers. 

\subsection{Drivers for Applying Self-Adaptation (RQ1)}
\label{subsec:rq1}
We now analyse the data that we collected for answering RQ1. This research question focuses on the drivers of practitioners for applying self-adaptation and the types of problems they solve using self-adaptation. Note that the data used to answer RQ1 comes from the 100 participants that have experience with concrete self-adaptive systems (i.e., the participants that answered ``Yes'' to Q0.1). 

\vspace{5pt}\subsubsection{For which problems do you apply self-adaptation? (Q1.1)}

\begin{figure}[h]
\centering
\includegraphics[width=0.85\columnwidth]{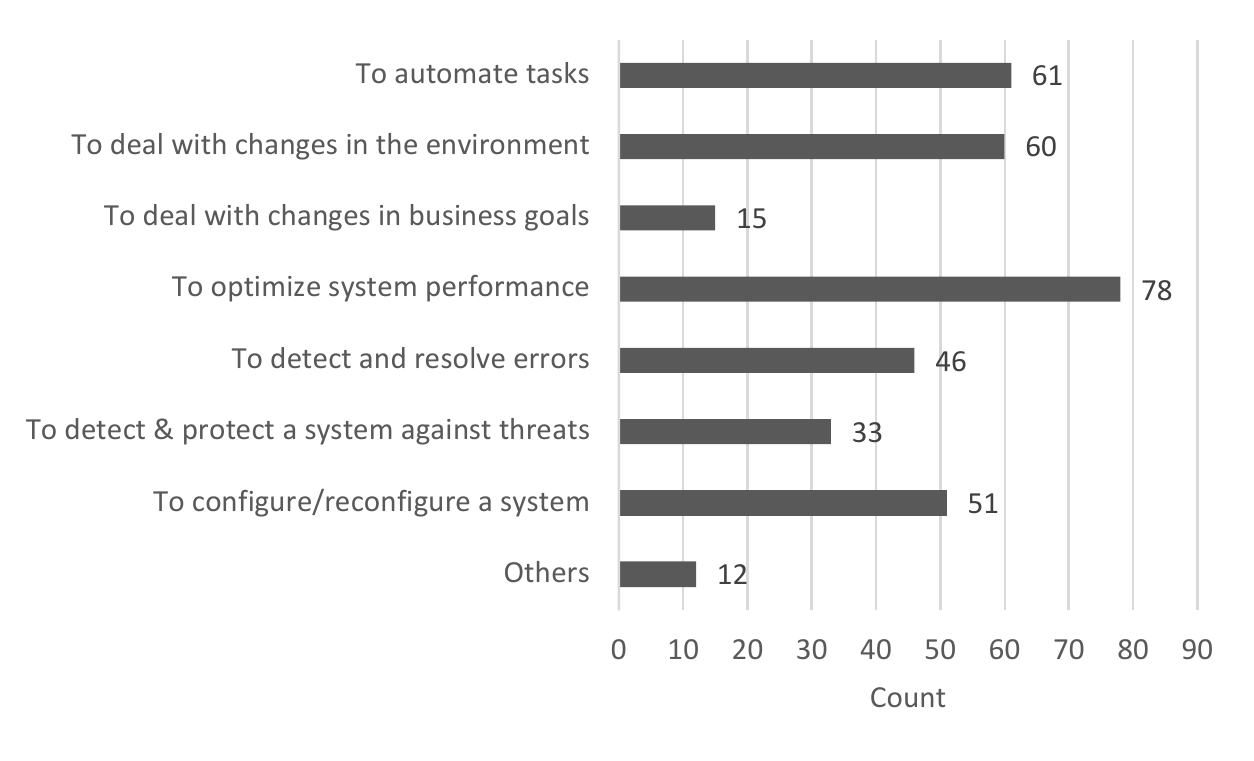}\vspace{-10pt}
\caption{Problems to apply self-adaptation (Q1.1).}
\label{fig:q1-1}
\end{figure}

Figure~\ref{fig:q1-1} summarises the results. On average, the participants applied self-adaptation for 3.6 types of problems from the predefined list (with seven options). The results show that practitioners apply self-adaptation to deal with a variety of problems. Optimising performance and automating tasks are the main problems addressed by self-adaptation in industry. On the other hand, dealing with changes in business goals is less frequently solved using self-adaptation. `Others' include for example support for testing and evolution. 

\vspace{5pt}\subsubsection{What are the main business motivations to apply self-adaptation? (Q1.2)}

\begin{figure}[h]
\centering
\includegraphics[width=0.65\columnwidth]{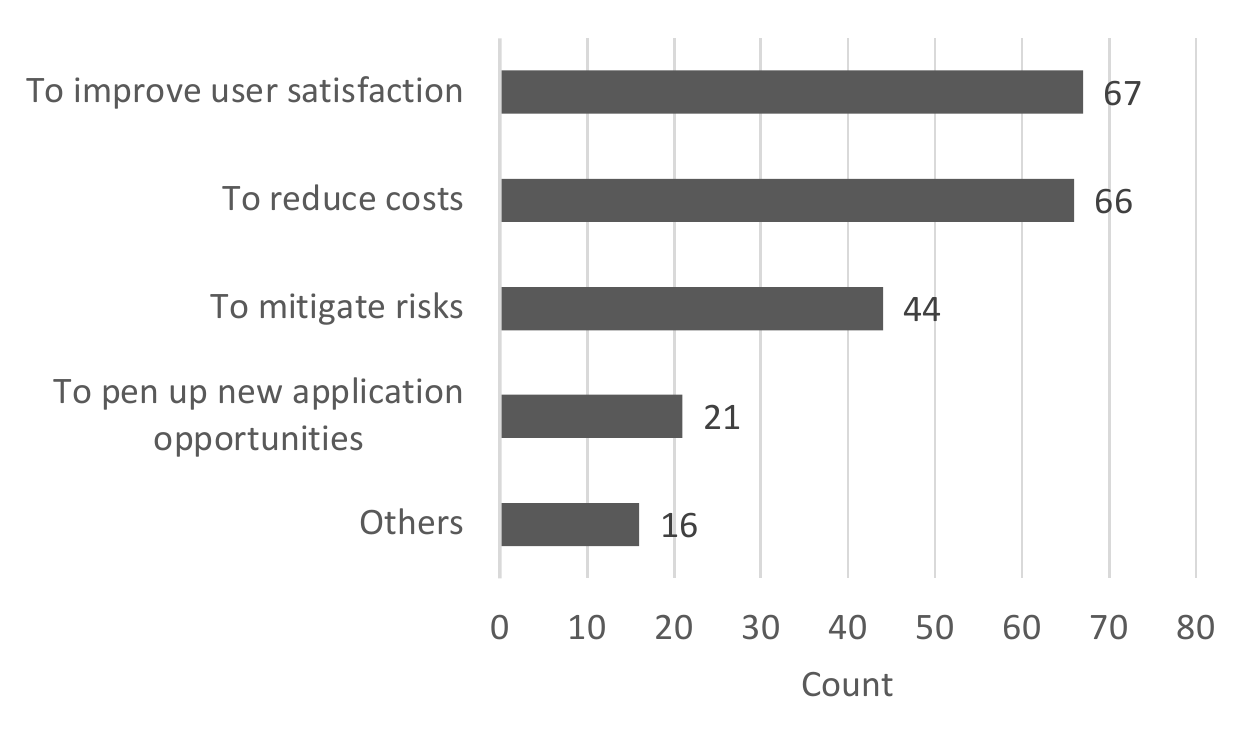}\vspace{-10pt}
\caption{Main business motivations to apply self-adaptation (Q1.2).}\vspace{-10pt}
\label{fig:q1-2}
\end{figure}

Figure~\ref{fig:q1-2} summarises the results. On average, the participants provided 2.1 business motivations to apply self-adaptation. Improving user satisfaction, reducing costs, and mitigating risks are the most selected motivations for using self-adaptation. Opening up new application opportunities was selected by a lower number of 21 participants. Examples of `Others' are improving utility and managing complexity. 

\vspace{5pt}\subsubsection{What could be benefits of applying self-adaptation? (Q1.3)}

Ninety-two participants 
provided meaningful descriptions of benefits of self-adaptation, an average of 1.8 benefits per participant. 

\textbf{Analysis of comments:} 
We summarise the findings in Table~\ref{tab:codes_q1-3}. For each category (bold font) and code, we include how often it appeared and we provide a few example quotes.\footnote{Categories have codes which express more concrete instances of the categories.}
The dominating benefits of applying self-adaptation are \textit{improved utility} (61 participants), \textit{savings} in costs and resources (38 participants), and \textit{improved human interaction} (37 participants).  


\small
\begin{table}[hbt]
\caption{Comments: Reported benefits of self-adaptation (Q1.3).}
\label{tab:codes_q1-3}
\begin{tabular}{p{3cm}lp{9.6cm}}
\hline\noalign{\smallskip}
Categories/codes & \# & Example quotes\\
\noalign{\smallskip}\hline\noalign{\smallskip}
\textbf{Improved utility} & \textbf{61} & \\
Robustness & 21 & ``fault tolerance, one node dies, a new one is spawned without manual intervention''; ``better error handling and prompt disaster recovery'' \\
Performance & 16 & ``Improve performance and quality-of-service''; ``increase in the speed of adaptation''  \\
Availability & 8 & ``The main benefit for us is the 99.9999\% availability, which is crucial for some customers of these cloud-specific solutions''\\
Other & 16 & ``for IoT: optimized operations, improved energy usage''; ``an important part to guarantee the safety [...] of the overall system.''\\\noalign{\smallskip}\hline\noalign{\smallskip}
\textbf{Savings} & \textbf{38} & \\
Costs & 25 & ``The primary benefit is cost reduction''; ``the cheaper bills for running this in an efficient manner in e.g. a cloud service''\\
Resources & 13 & ``scales down resources during hours when traffic is low, and scales up during peak hours, without any manual interference.''\\\noalign{\smallskip}\hline\noalign{\smallskip}
\textbf{Improved human interaction}  & \textbf{37} & \\
User experience & 19 & ``Keep Telco network in optimal condition so that QoS and user experience is maximized, and churn minimized''; ``better user satisfaction because of prompt website responses''\\
Engineers support & 18 & ``removes most of the optimization burden from programmers, so they can be more productive''; ``Reduce workload on human operators; make (the results of) certain actions [...] repeatable and predictable''\\\noalign{\smallskip}\hline\noalign{\smallskip}
\textbf{Handle dynamics} & \textbf{22} & \\
Load dynamics & 12 & ``Change AGV behavior depending of the workload with the goal to save energy (battery life).''\\
Context dynamics & 10 & ``Each machine is unique and its optimal operational parameters change over time due to ware, location, task and seasonal factor.''\\\noalign{\smallskip}\hline\noalign{\smallskip}
\textbf{Other improvements}
& \textbf{16} & \\
Various & 16 & ``In case of spikes in incoming events the system is able to adapt [...] avoiding bottlenecks.''; ``Easier and faster market integration''; ``It's fundamental in huge infrastructure systems otherwise we can’t make it happen.''\\
\noalign{\smallskip}\hline
\end{tabular}
\end{table}
\normalsize

\begin{framed}
\noindent \textbf{Key insight(s) from RQ1:} 
\begin{enumerate}
\item Self-adaptation is widely applied in industry across a wide variety of domains. 
\item Practitioners primarily apply self-adaptation to optimise performance, automate tasks, and deal with changes in the deployment environment. 
\item The dominating business motives to apply self-adaptation in industry are primarily improving user satisfaction and reducing costs, and secondarily 
mitigating risks. 
\item The main benefits of applying self-adaptation are improved utility (in robustness and performance), savings (costs and resources), improved human interaction (user experience and engineers support), and handling dynamics (in the context and system load).  
\end{enumerate}
\end{framed}

\subsection{RQ2: Characterisation of Self-adaptation}
\label{subsec:rq2}

\subsubsection{Explain a concrete self-adaptive system you worked with (Q2.1)}

Except for one, all participants with experience in self-adaptation provided a concrete description of a system they worked with. 

\textbf{Analysis of comments:} Tables~\ref{tab:codes_q2-1.I} and~\ref{tab:codes_q2-1.II} summarise the findings. We focused on characteristics of self-adaptive systems and identified three categories: \emph{subject}, \emph{type}, and \emph{trigger} of adaptation. With subject we mean the system or part of it that is adapted; type refers to the kind of adaptation that is applied, and trigger refers to the source that initiates adaptation. 

\small
\begin{table*}[b!]
\caption{Analysis of comments I -- Explain a concrete self-adaptive system you worked with (Q2.1)}
\label{tab:codes_q2-1.I}
\begin{tabular}{p{3cm}lp{9.6cm}}
\hline\noalign{\smallskip}
Categories and codes & \# & Example quotes\\
\noalign{\smallskip}\hline\noalign{\smallskip}
\textbf{Subject of adaptation} & \textbf{99} &  \\

System & 28 & ``Our company develops safety critical systems for railway. Systems architecture is often with redundancy - e.g. 2 out of 3 system, where is automatic reconfiguration implemented. Purpose is high safety and availability''; ``A flexible manufacturing system ... the system and the individual station within the system can "sense" what kind of work piece it has in front of itself and what it or another machine should do with it in the next step.'' 
\\

Module & 22 & ``Environment compensation system for capacitive touch interface. Such system is influenced by envirenmental change (for example temperature)''; ``We manage the memory usage of the process. Once memory usage over a limit (i.e. 90\%), we start throttling the workload.''\\

Platform layer & 13 & ``Monitoring the memory/CPU/disk consumption of our servers and suggesting measures to fix it through human intervention.'' 
\\

Application layer & 11 & ``HotSpot JVM ... reads a program's Java bytecode, and adaptively tunes the performance of the program at runtime, adapting to runtime profiles.''\\

Cluster & 10 & ``Spark executor auto-scaling system. We built this system to automatically add or remove nodes to our Spark cluster when we have a high demand of resources from our Spark jobs.'' \\

Network & 6 & 
``"Our radios apply 'channel assessment' ... that optimizes the radio channels used during BLE communication. Our radios also apply very aggressive power management. peripherals and cores are switched off whenever possible to minimize the system's power usage."'' 
\\

Mixed & 6 & ``Enterprise-cloud environment consisting of dozens of different (micro) services providing functionality to 3rd parties as well as internal employees - data management, authentication and authorization, business process automation, as well as internal development process support (build servers, logging, etc.).''
\\

CI/CD pipeline & 3 & ``Sacling up and down our infrastructure (CI/CD) chain to build and integrate the truck software.'' \\
\noalign{\smallskip}\hline
\end{tabular}
\end{table*}

\begin{table*}[h!]
\caption{Analysis of comments II -- Explain a concrete self-adaptive system you worked with (Q2.1)}
\label{tab:codes_q2-1.II}
\begin{tabular}{p{3cm}lp{9.6cm}}
\hline\noalign{\smallskip}
Categories and codes & \# & Example quotes\\
\noalign{\smallskip}\hline\noalign{\smallskip}
\textbf{Type of adaptation} & \textbf{99} &  \\

Auto-scaling & 33 & ``Automated horizontal scaling of AWS EC2 instances for medical data processing systems''; ``autoscale a cluster based on the resource usage of the nodes of the cluster.''\\
Auto-tuning & 28 & ``A mink feeding robot, that can adjust the food amount according to a set of feeding rules and the food left over from last feeding.'' \\
Monitor/Analysis & 22 & ``We configured AWS alarms to monitor performance of our systems in case we get more than few number of HTTP 400/500 errors''; ``Monitoring the memory/CPU/disk consumption of our servers and suggesting measures to fix it through human intervention.'' \\
Automated reconfiguration & 11 & ``Continuos integration system - Other \& starts building \& testing a new version as soon as it detects code changes  Build alignment - Creates a new release whenever a subsystem builds successfully.''\\
Other & 5 & ``Our mobile robots scan their environments using laser scanners and other sensors and plan their behavior accordingly.'' ``self healing automotive systems''\\\noalign{\smallskip}\hline\noalign{\smallskip}
\textbf{Trigger for adaptation} & \textbf{99} &  \\
System properties & 27 & ``Auto-scaling functionality of an Azure Service Fabric cluster running a transformation load for processing AGV statistical and playback data.''; ``Realtime focused data streaming protocol ... must take care to avoid exhausting the network resources and thus incurring packet loss and latency spikes, which are very noticeable in games.''\\

Environment properties & 18 & ``An IoT system running in Kubernetes and used to monitor water leaking for household insurance.''; ``A flexible manufacturing system ... can "sense" what kind of work piece it has in front of itself and what it or another machine should do with it in the next step.'' \\

System load & 14 & ``Kubernetes, for handling load intensive periods for scaling up, and self recover from crashes.''; ``Autoscaling of SaaS applications in function of load on AWS and Azure clouds.'' \\

Events & 12 & ``We use kubernetes which provides notification callbacks on any event such as host/pod not available, based on these events we auto mark the node was inactive and do not use those nodes for further write or read operations''; ``Auto Scaling an EMR cluster in AWS based on incoming event data''\\

User actions & 7 & ``[adapt] cache warm up strategy based on user interactions''; ``scammers ... To decide the users that are most likely to be a scammer, the system tracks the past performance of models responsible for flagging potential scammers.''  \\
\noalign{\smallskip}\hline
\end{tabular}
\end{table*}
\normalsize

Ninety-nine participants provided a description of what is the subject of adaptation in the systems they work with. Top results are \emph{system} that occurred 28 times, followed by \emph{module} with 22 times (i.e., a part of a system). \emph{Platform layer} (infrastructure, execution platform, etc.) was mentioned 11 times and \emph{application layer} 11 times. 

Of the participants that worked with self-adaptation, 86 described in total 101 instances of the types of adaptation they apply (i.e., an average of 1.17). \emph{Auto-scaling} with 33 occurrences and \emph{auto-tuning} with 28 are the most frequent types of adaptations applied by the participants. Twenty-two participants focus on \emph{monitoring and analysis} only (they may use the human in the loop for other adaptation functions). 

Finally, 62 participants explained in total 78 triggers of adaptation in their work (i.e., an average of 1.21 triggers). The main triggers originate from \emph{system properties} with 27 occurrences and \emph{environment properties} with 18 occurrences. Changes in the system load, events,\footnote{An event is an occurrence or action that happens asynchronously at some point in time, such as an alarm, an alert, etc.} and user actions are the other types of triggers for adaptation.

\begin{framed}
\noindent \textbf{Key insight(s) from RQ2:} 
\begin{enumerate}
\item Self-adaptation is applied at different levels of industrial software-intensive systems: from a complete system to parts of a system and support systems.  
\item The dominating types of adaptations applied in industry are auto-scaling, auto-tuning, and monitoring/analysis. 
\item Adaptions in industrial software-intensive systems are triggered by changes in properties of systems and their environments, dynamics in system load, relevant events, and through user actions.  
\item Technologies such as elastic cloud and  auto-scalers are key enablers for the realisation of self-adaptation in practice.   

\end{enumerate}
\end{framed}

\subsection{RQ3: Application of Self-adaptation}
\label{subsec:rq3}

\subsubsection{What mechanisms or tools does the self-adaptive system you worked with use to monitor a managed system during operation? (Q3.1)}

The participants provided a total of 146 instances of mechanisms or tools they used for monitoring in a self-adaptive system they worked with, i.e., on average, 1.5 mechanisms/tools per participant. 

\textbf{Analysis of comments:} Table~\ref{tab:codes_q3-1} summarises the findings. The participants focused on both ``what'' is being monitored and ``how'' monitoring is done. Based on this we identified three categories: \emph{monitoring metrics}, \emph{monitoring mechanisms}, and \emph{monitoring tools}. Of the 100 answers, we marked 14 as unclear. 

\small
\begin{table*}[t!]
\caption{Analysis of comments -- Mechanisms or tools used to monitor a managed system (Q3.1).}
\label{tab:codes_q3-1}
\begin{tabular}{p{3cm}lp{9.6cm}}
\hline\noalign{\smallskip}
Categories and codes & \# & Example quotes\\
\noalign{\smallskip}\hline\noalign{\smallskip}

\textbf{Monitoring metric} & \textbf{75} &  \\

Resource usage & 23 & ``Active sessions counting, resource utilisation (e.g. RAM) monitoring given by VM''; ``Typically CPU and Memory usage''; ``Helsim: uses CPU counters to measure time or power consumption to process particles''\\

Load & 18 & ``Number of incoming HTTP requests''; ``The system polls the queue of the Spark job scheduler in our cluster every 5 seconds via REST API, using a NiFi flow.''; ``Number of queries''; ``number of requests''\\

Reliability metrics & 13 & ``AWS lambda error metric is monitored to see if the sum of 400/500 errors for every part 5 mins is less than some specified amount.''
\\

Performance metrics & 12 & ``We track the response times for the users' requests.''; ``monitored systems implement specific features to provide data about their performance.'' \\

Application state & 9 & ``Tracking properties are - correct integrity - functionality of memorries (RAM, ROM), correct values and integrity of data among redundant parts.'' 
\\\noalign{\smallskip}\hline\noalign{\smallskip}
\textbf{Monitoring mechanism} & \textbf{20} &  \\
Environment sensors & 9 & ``Based on external information (external sensors like Lidar, Camera, GPS, ...) making sure no accident were to happen''; ``Exteroceptive are aggregated to create a snapshot of the world's state. These are LIDAR and Image sensors. We use Proprioceptive sensors to determine the robot's state. These are encoders only.'' \\

Logging mechanisms & 6 & ``Logging software triggered whenever an incoming request is made''; ``The system logs all interactions, both errors and successful operations.''\\

System sensors & 4 & ``Based on internal information (internal sensors like Wheel speed, steering angle, yaw and roll sensors, ...) optimize the performance to support the driver to drive optimal.''\\

Humans & 1 & ``Human review decisions are used to monitor the precision of models.''
\\\noalign{\smallskip}\hline\noalign{\smallskip}
\textbf{Monitoring tool} & \textbf{34} &  \\

Kubernetes monitoring & 9 & ``Kubernetes clusters are made out of master and worker machine nodes. On the worker nodes runs a process called kubelet that monitors the state of the worker nodes in the Kubernetes cluster''; ``Probes implemented in the application, metrics provided by K8s metrics server (goes down to cgroups via kubelet)'' \\

Prometheus & 9 & ``- every service exposes a defined set of metrics. We collect metrics regarding every layer of the distributed system. We mainly use Prometheus and Splunk to collect these metrics.''; ``Prometheus and grafana for monitoring health of services'' \\

AWS monitoring & 8 & ``We use AWS CloudWatch service to monitor and act on any event with ServerLess AWS lambda functions.''; ``AWS Lambda based monitor which monitor aprox number of message in SQS queue'' \\

Other: Azure monitoring, Datadog, Splunk, cAdvisor, Elasticsearch & 8 & ``Default tooling from Azure / AWS in combination with splunk''; ``We are using Datadog to collect relevant metrics.''; ``AKS monitors the system load and response time to start-up more instances. It also checks for malfunctioning applications and restarts them when stalled, providing high availability.'' \\

\noalign{\smallskip}\hline
\end{tabular}
\end{table*}
\normalsize

The participants mentioned in total 75 metrics they use for monitoring. \emph{Resource usage} with 23 occurrences, \emph{system load} with 18, and \emph{reliability} with 13 are the most frequently mentioned metrics. 

The participants described in total 20 monitoring mechanisms. \emph{Environment sensors} occurred nine times and \emph{system sensors} four times. Six participants described different \emph{logging mechanisms}, and in one system, a \emph{human} is involved in monitoring. 

Finally, the participants provided in total 34 tools they use for monitoring. The most prominent tools are \emph{Kubernetes monitoring} and \emph{Prometheus}, which each occurred nine times, followed by \emph{AWS monitoring} with eight occurrences.\footnote{https://kubernetes.io/ - https://prometheus.io/ - https://aws.amazon.com/}

\subsubsection{What mechanisms or tools does the self-adaptive system you worked with use analyse conditions of a managed system during operation? (Q3.2)}

The participants provided a total of 115 instances of mechanisms or tools they used for analysing conditions of a self-adaptive system they worked with, i.e., on average 1.5 mechanisms/tools per participant.

\textbf{Analysis of comments:} Table~\ref{tab:codes_q3-2} summarises the findings.
We identified two categories: \emph{analysis mechanisms} and \emph{analysis tools}. Out of the 100 valid answers, 21 were marked as unclear or not applicable (such as ''Fairy simple algoritms'' or ''The tech stack we use is proprietary and the tools we use are built in house''). 
The rest of the participants mentioned in total 73 mechanisms they use for analysis. 
The most frequently mentioned mechanisms are \emph{data analysis methods} (such as interference, statistical data analysis, what-if analysis, and search-based methods) with 18 occurrences, \emph{comparison to threshold} with 16 occurrences, and \emph{metric(s) calculation} and \emph{learning} (mostly machine learning) with 12 occurrences. 
The participants provided in total 23 tools they use for analysis.
\emph{AWS analysis tools} occurred nine times, followed by \emph{Kubernetes stack} with seven, and \emph{Dynatrace} with two occurrences. 
Other tools mentioned by the participants include Splunk, JMX, Jasmina, Azure, Openshift, and Kibana.

\small
\begin{table*}[h!]
\caption{Analysis comments -- Mechanisms or tools used to analyze conditions of a managed system (Q3.2).}
\label{tab:codes_q3-2}
\begin{tabular}{p{3cm}lp{9.6cm}}
\hline\noalign{\smallskip}
Categories and codes & \# & Example quotes\\
\noalign{\smallskip}\hline\noalign{\smallskip}

\textbf{Analysis mechanism}  & \textbf{73} &  \\

Data analysis methods & 18 & ``I think it uses some rolling average or some similar algorithm to estimate whether to scale up or down.''; ``simple statistical inferences based on metrics and simple rules encoded by developers.''; ``statistical analysis of data''\\

Comparison to threshold & 16 & ``Comparing the error rate with constant/dynamic thresholds.''; ``Hard  coded critical boundaries like min max values which lead to switching over to emergency modes [...]''; ``when it falls below Service Level Agreements this indicates a need for auto-scaling'' \\

Metric(s) calculation & 12 & ``Failure rate is used to measure quality of adaptation parameters.''; ``Capturing performance of each node. ''; ``Measurement of traffic load, CPU utilization, and general availability metrics (reachability, status, ...)'' \\

Learning & 12 & ``Each station has a kind of edge computing component that performs some analysis based on machine learning results.''; ``It tracks both the internal working conditions (load) of itself as a serving component, and learns about overall serving conditions.''; ``The system uses biosensory feedback to determine the riders' happiness [...]'' \\

Custom rules & 9 & ``Mostly a simple ruleset gleaned by experimentation and observing how the resulting adaption steps perform at runtime.''; ``we have alertmanager to set up some rules that are known to be issues that have clear solutions'' \\

Autoscaling policy & 5 & ``[...] the response of the scheduler is parsed and the queue length is evaluated. If greater than zero, the flow performs a SCALE UP operation. If equal to zero, the flow performs a SCALE DOWN operation.'' \\

Semantic reasoning & 1 & ``Reasoning on knowledge graphs'' 
\\\noalign{\smallskip}\hline\noalign{\smallskip}
\textbf{Analysis tool} & \textbf{23}
\\
AWS analysis tools & 9 & ``Analytics functions native to the cloud environment the system runs in (AWS).''; ``AWS based auto-scaling conditions as provided in the Cloud formation setup of the cluster'' \\

Kubernetes stack & 7 & ``The master nodes have all sorts of different components such as the kube-scheduler, controllers and state db (etcd), that are managed via the kube-apiserver. ''; ``Built-in Kubernetes/Openshift mechanisms [...]'' \\

Dynatrace & 2 & ``analyze was done by Dynatrace or by Keptn itself by checking against thresholds'' \\

Other & 5 & ``We mainly use rule-based systems like Splunk to automatically analyse production metrics against patterns.''; ``Default tooling from Azure''; ``Kibana'' \\

\noalign{\smallskip}\hline
\end{tabular}
\end{table*}
\normalsize

\subsubsection{What mechanisms or tools does the self-adaptive system you worked with use to change a managed system or parts of it during operation? (Q3.3)}
The participants provided 126 instances of mechanisms or tools they have used for applying changes, i.e., 1.3 mechanism/tool per participant.

\textbf{Analysis of comments:} Table~\ref{tab:codes_q3-3} summarises the findings.
Out of the 100 valid answers, 23 were marked as unclear or not applicable. We identified two categories: \emph{change mechanisms} and \emph{change enacting tools}. In total, 83 instances of mechanisms for change were reported. \emph{Scaling mechanisms} with 36 occurrences and \emph{reconfiguration} (changing the adaptation logic, network reconfiguration, parameter adjusting, load balancing) with 25 occurrences are the most frequently mentioned changing mechanisms. Twelve participants used \emph{non-automated mechanisms} that refer to notifications and change tasks done by humans. 
The participants mentioned 19 tools they use for enacting change. \emph{Kubernetes} occurred nine times, \emph{AWS} seven times and other tools, including Openshift and Dynatrace, three times. 

\small
\begin{table*}[hbt]
\caption{Analysis of comments -- Mechanisms or tools used to change a managed system or parts of it (Q3.3).}
\label{tab:codes_q3-3}
\begin{tabular}{p{3cm}lp{9.7cm}}
\hline\noalign{\smallskip}
Categories and codes & \# & Example quotes\\
\noalign{\smallskip}\hline\noalign{\smallskip}

\textbf{Change mechanism} & \textbf{83} &  \\

Scaling mechanisms & 36 & ``The server-side system has a load balancer. Hence we increase the number of workers behind the load balancer to decrease the average response time for the users.''; ``It adjusts the number of worker nodes.''; ``Adding a completely similar server / serverless Lambda instance''; \\


Reconfiguration & 25 & ``The adaptation directly adjusts the period between the packet send events, as well as the number of packets allowed during each send event. [...]; ``Depending on context, controlled variables are managed through different automation systems.''; ``reconfiguration of the management entity ... to support a larger (or smaller) scale distributed system''; ``load balancer/director that may support controlling the exposure facade towards the system environment. '' \\



Non-automated & 12 & ``To effect change on the managed system, the results from the tool need to be approved by an engineer, and are then acted on by the mining and plant teams. These processes are for the most part not automated [...].''; ``Generating alerts and expecting humans to resolve the error manually based on suggestions.''; ``Did not do this [...]. Based on safety protocols this could not be secured'' \\




Restarting/deploying & 7 & ``Mostly just restarting the managed subsystems. In the case of Kubernetes HPA, its the horizontal scaling (up/down) of the Pods''; ``Generally restarts the unhealthy workload, but in the case of autoscaling can also be used to add or remove replicas''; ``... our pipelines use simple bash scripts to deploy previous versions when new versions fail.'' \\


Migration & 3 & ``Once the control process informs the control plane, it starts a workflow what we call as instance warming workflow which will dump items that supposed to go to that node from another replica and fills them.''; ``virtual machine (VM) migration or creation.'' 
%
%
\\\noalign{\smallskip}\hline\noalign{\smallskip}
\textbf{Change enacting tool} & \textbf{19}
\\
Kubernetes & 9 & ``Mostly just restarting the managed subsystems. In the case of Kubernetes HPA, its the horizontal scaling (up/down) of the Pods''; ``... to change topology we simply use K8S api to add/remove worker pods'' \\

AWS & 7 & ``AWS based in-built auto scaling capabilities ''; ``Use the AWS ElasticLoadBalancer and also trigger actions via AWS Lamda functions when required.'' \\

Other & 3 & ``IBM ITM, Log Analyzer, TCAM''; ``UC4 Automation Engine workflows that orchestrate kubernetes clusters''; ``Build-in Openshift mechanisms'' \\

\noalign{\smallskip}\hline
\end{tabular}
\end{table*}
\normalsize

\subsubsection{What is the degree of automation of the majority of the self-adaptive solutions you work with in your organization? (Q3.4)}

\begin{figure}[h]
\centering
\includegraphics[width=0.6\columnwidth]{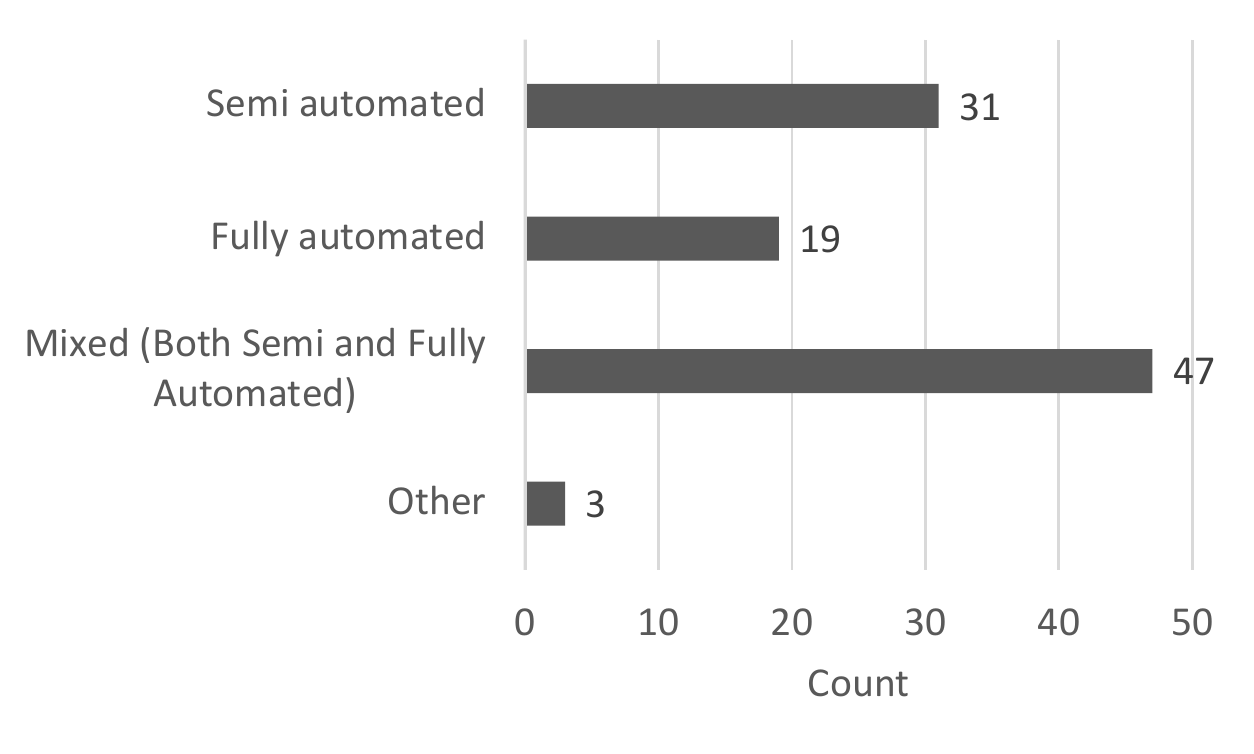}
\vspace{-10pt}
\caption{Degree of automation of the self-adaptive solutions the participant has worked with (Q3.4).}
\vspace{-10pt}
\label{fig:q3-4}
\end{figure}

All 100 participants provided an answer to this question; Figure~\ref{fig:q3-4} summarises the findings.
Forty-seven participants reported mixed automation in their projects (both semi and fully automated), while 31 indicated semi automation and 19 indicated full automation. Three participants selected other; two of them mentioned that there is no automation, the third stated ``fully-automated till first incident.''

\subsubsection{Do you reuse solutions to realise self-adaptation in systems you work with? (Q3.5)}
\begin{figure}[h]
\centering
\includegraphics[width=0.7\columnwidth]{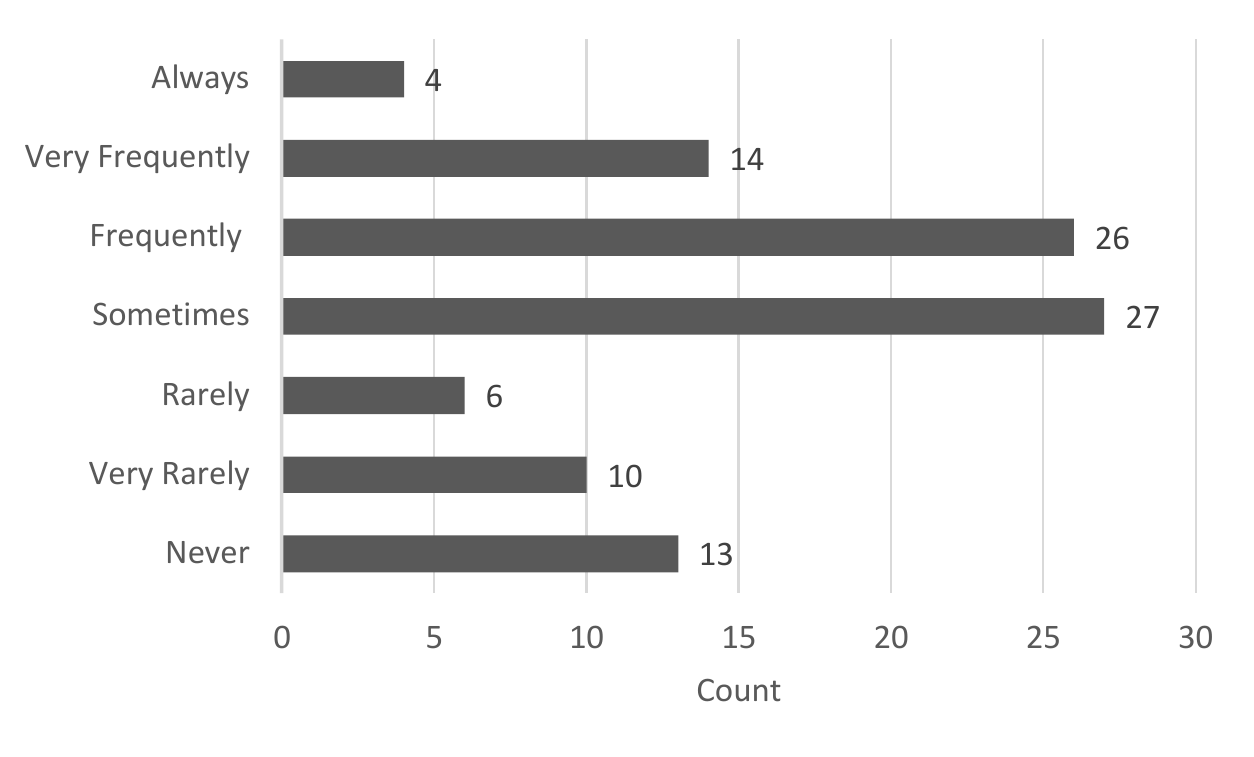}
\vspace{-10pt}
\caption{Do you reuse solutions to realise self-adaptation? (Q3.5)}
\vspace{-10pt}
\label{fig:q3-5}
\end{figure}

All 100 participants provided answers to this question that are summarised in Figure~\ref{fig:q3-5}.
A majority of 71 participants reuse at least sometimes solutions in self-adaptive systems (44 of them reuse solutions frequently to always). The other 29 participants rarely, very rarely or never reuse solutions. 

\subsubsection{Please provide a concrete example of reuse you used to realise self-adaptation? (Q3.6)}

Sixty-seven participants provided examples of reuse in the realisation of the self-adaptive systems. 

\textbf{Analysis of comments:} 
Table~\ref{tab:codes_q3-6} summarises the findings. We focused on the subjects of reuse and identified five categories: \emph{code}, \emph{design artifacts}, \emph{specifications}, \emph{IT infrastructure}, and \emph{procedures}. The 67 participants provided in total 91 objects of reuse in adaptation, i.e., an average of 1.4. Code occurred 33 times, with \emph{modules} as the top subject of reuse (18 instances). Design artifacts was mentioned 22 times with \emph{patterns} and \emph{architecture} as main subjects of reuse (each with seven instances). Specification was mentioned 18 times as objects of reuse, IT infrastructure 11 times, and procedures seven times. The results demonstrate that reuse in self-adaptation is common practice, although the use of patterns (a topic that gets increasing attention in research) is limited.  

\small
\begin{table}[hbt]
\caption{Comments: Examples of reuse in self-adaptive systems (Q3.6).}
\label{tab:codes_q3-6}
\begin{tabular}{p{3cm}lp{9.7cm}}
\hline\noalign{\smallskip}
Categories/codes & \# & Example quotes\\
\noalign{\smallskip}\hline\noalign{\smallskip}
\textbf{Code} & \textbf{33} &  \\
Modules & 18 & ``Self adaptation mechanisms used for speech recognition ... are also used for computer assisted coding solutions. ''; ``Different parts of the Behavior tree can be reused in different robots.''\\
Scripts and algorithms & 8 & ``The same scripts and solutions are constantly reused - because it’s the easiest way to create new with a constant lack of time.''; ``Threshold algorithms are reused frequently, with the threshold value adapted for the specific use case.''  \\
Libraries & 7 & ``internal libraries that simplify monitoring, interaction with external tools, etc''
\\\noalign{\smallskip}\hline\noalign{\smallskip}
\textbf{Design artifacts} & \textbf{22}
\\
Patterns & 7 & ``We try to reuse design patterns (e.g. autoscaling) for all cloud native applications we build.''; ``Re-use of design patterns like MAPE-K. ''\\
Architecture & 7 & ``AWS stack\,...\,can be used as a generic template cross different applications which are based on a job processing ''\\
Know-how & 5 & ``We use similar principles in different product.''; ``We reused knowledge of driver parameter adaptation from FDM (3 axis) printer while designing a SLA (single axis) printer.''\\
Models & 3 & ``machine learning cost models can be reused by different systems''
\\\noalign{\smallskip}\hline\noalign{\smallskip}
\textbf{Specifications} & \textbf{18}
\\
Policies \& rules & 5 & ``auto-scaling policies ... have a standard definition which can be reused in different systems or use-cases.''\\
Configuration files & 5 & ``K8s config files for different cloud native application can be similar''\\
Templates & 4 & ``We reuse very similar set of configuration templates of container deployment''\\
Metrics & 4 & ``Kibana alerts''
\\\noalign{\smallskip}\hline\noalign{\smallskip}
\textbf{IT infrastructure} & \textbf{11}
\\
Frameworks\,\&\,platforms & 7 & ``a framework for monitoring metrics that allows labels to be given to properties, the time-series data to be tracked in a database, and then hooks to visualization database and alert systems.''\\
Tools & 4 & ``Use the same tools AWS provides for all our different product deployments.''
\\\noalign{\smallskip}\hline\noalign{\smallskip}
\textbf{Procedures} & \textbf{7}
\\
Processes & 3 & ``Writing "watchdog" processes for systems that aren't deployed to kubernetes''\\
Pipelines & 2 & ``pipeline (Application\,-\,Datadog\,-\,custom logic\,-\,AWS API) is replicated with different settings for different use-cases.''\\
Schedules & 2 & ``Most of the approaches we use for digital twins share some history ... An example of that is in the scheduling space, where schedules need to adapt to changes in resources or the inclusion and removal of tasks.''\\
\noalign{\smallskip}\hline
\end{tabular}
\end{table}
\normalsize

\subsubsection{Why do you not often reuse solutions
when realising self-adaptive systems?
What hinders their reuse, please provide a
short answer? (Q3.7)}

This was a conditional question that was only asked to the participants that answered never or very rarely to Q3.5 (that asked whether participants reuse solutions to realise self-adaptation). Twenty-three participants provided such an answer to Q3.5.  

\textbf{Analysis of comments:} 
Table~\ref{tab:codes_q3-7} summarises the findings. From 18 participants that provided valid answers, we identified 19 \emph{reuse hurdles}, i.e., an average of 1.1. The main hurdle reported by 11 participants is \emph{difference in problems}, hampering easy reuse. Other hurdles are \emph{lack of experience or maturity} in applying self-adaptation within the company (4 occurrences), and the \emph{complexity of the system} and \emph{organisational concerns} (each with 2 occurrences). 

\small
\begin{table}[hbt]
\caption{Comments: Why not often reusing solutions
when realising self-adaptive systems (Q3.7).}
\label{tab:codes_q3-7}
\begin{tabular}{p{3.6cm}lp{8.8cm}}
\hline\noalign{\smallskip}
Categories/codes & \# & Example quotes\\
\noalign{\smallskip}\hline\noalign{\smallskip}
\textbf{Reuse hurdles} & \textbf{19} &  \\
Different problems & 11 & ``In my case every self-tuning problem is different and prevents easy reuse.''; ``Our applications and application domains are very different and since we do research we actively look for new and different challenges.''\\
Lack of experience/maturity & 4 & ``I think lack of competence is a huge thing to overcome, though most of the organisations around us try to catch up '' \\
System structure & 2 & ``The solutions were too coupled, too integrated and not enough modularized.''\\
Organisational concerns & 2 & ``We have to go through a legal department in order to reuse code from outside ... That poses a large problem. ''\\
\noalign{\smallskip}\hline
\end{tabular}
\end{table}
\normalsize

\subsubsection{How do you ensure that you can trust the self-adaptive solutions you build? (Q3.8)}

Ninety-one of the 100 participants that worked with self-adaptation provided valid answers. 

\textbf{Analysis of comments:} 
Table~\ref{tab:codes_q3-8} summarises the findings. The participants provided in total 152 instances of techniques for ensuring trust in the self-adaptive systems they build, i.e., on average 1.7 techniques per participant. We grouped the techniques in three categories: \emph{testing and verification}, \emph{stakeholder-centred techniques}, \emph{online techniques}. 
Testing and verification was mentioned 71 times with \emph{extensive testing} being the main technique occurring 58 times, followed by \emph{benchmarking} occurring 10 times and verification (three times). Stakeholder-centred techniques were mentioned 45 times. In this category, \emph{human supervision} (22 occurrences) and \emph{rigorous design and development} (10 occurrences) were the main reported techniques. Finally, online techniques were mentioned 36 times with \emph{runtime monitoring and alerting} as main reported technique (27 occurrences). In contrast to an important focus of research in self-adaptation, (formal) \emph{verification} as a technique to ensure trust was only mentioned three times.   

\small
\begin{table*}[hbt]
\caption{Analysis of comments - Techniques for ensuring trust in self-adaptive solutions (Q3.8).}
\label{tab:codes_q3-8}
\begin{tabular}{p{3.6cm}lp{8.8cm}}
\hline\noalign{\smallskip}
Categories and codes & \# & Example quotes\\
\noalign{\smallskip}\hline\noalign{\smallskip}

\textbf{Testing and verification}  & \textbf{71} &  \\

Extensive testing & 58 & ``We use extensive testing (unit, module, system)''; ``We have extensive testing on test k8s clusters, provisioned for these purposes. ''; ``We have countless amount of testing and verification code built as part of the OpenJDK to ensure the quality of the product is appropriate. ''\\

Benchmarking & 10 & `As a lot of the self adaptation logic involves optimization opportunities, we also regularly run many benchmarks and immediately report regressions''; ``We do testing of the machine learing models, but we also have pilot factories where we test our methods and design to see if all station perform as itended.'' \\

Verification & 3 & ``expert testing, supervision, verification when applicable''; ``Testing, but also some human verification as part of the Cloud Operations team.''
\\
\noalign{\smallskip}\hline\noalign{\smallskip}
\textbf{Stakeholder-centred techniques} & \textbf{45} &  \\

Human supervision & 22 & ``Human supervision until confident.''; ``Extensive system testing and gradual release of human supervision levels upon system going live.'' \\

Rigorous design and development & 10 & ``''; ``virtual training to ensure operators understand and are comfortable with the conditions in which the safety system will engage.'' \\

Trust in third-party software & 8 & ``for features like auto-scaling compute ... we use trusted vendors and deploy these features mainly for analytics use cases which are not business-critical.'' \\

Operational constraints & 5 & ``the concrete actions that are taken by the system are defined by the user. so there is never a surprise. the system only decides if and when to apply these actions.''; ``Our autotuning algorithms never fail for particular (exactly specified) set of systems. If the system fulfils these assumptions, it works always.'' 
\\
\noalign{\smallskip}\hline\noalign{\smallskip}
\textbf{Online techniques} & \textbf{36} &  \\

Runtime monitoring and alerting & 27 & ``In cases where an existing system is not being replaced but rather new capability is being added, results will be tracked over time to ensure accuracy.''; ``we have deployed some alert to track the high-level properties of the system.'' \\

Continuous testing during operation & 6 & ``there is gradual canary testing in the real production system. ''; ``Automated test scripts, automated "synthetic transactions" in production, model performance validation'' \\

Mitigation strategies & 3 & ``This automation can provide alter with all the steps and rollback automatically if there is any issue. '' \\

\noalign{\smallskip}\hline
\end{tabular}
\end{table*}
\normalsize

\begin{framed}
\noindent \textbf{Key insight(s) from RQ3:} 
\begin{enumerate}
\item Resource usage and system load are the main types of monitoring metrics used in practice. These metrics are primarily tracked by sensors in the environment and the system.  
\item Practitioners use various mechanisms for analysis in realising self-adaptation, with data analysis methods and comparison to thresholds as main mechanisms.   
\item A wide range of mechanisms are used to enact self-adaptation in industrial systems with auto-scaling and reconfiguration as top mechanisms. 
\item Practitioners extensively rely on tools such as Kubernetes and AWS to support the realisation of different functions of self-adaptation. 
\item Industrial systems apply a mix of semi and fully automated adaptation. 
\item A majority of practitioners reuse solutions when applying self-adaptation, mainly in the form of code, design artifacts, and specifications. 
\item Ensuring trust in industrial self-adaptive systems is mainly achieved through extensive testing, runtime monitoring and alerting, and human supervision. 
\end{enumerate}
\end{framed}

\subsection{RQ4: Difficulties, Problem Support, and Opportunities}
\label{subsec:rq4}

\subsubsection{Did you encounter particular difficulties when engineering or maintaining self-adaptive systems you worked with? (Q4.1)} 
\begin{figure}[h]
\centering
\includegraphics[width=0.7\columnwidth]{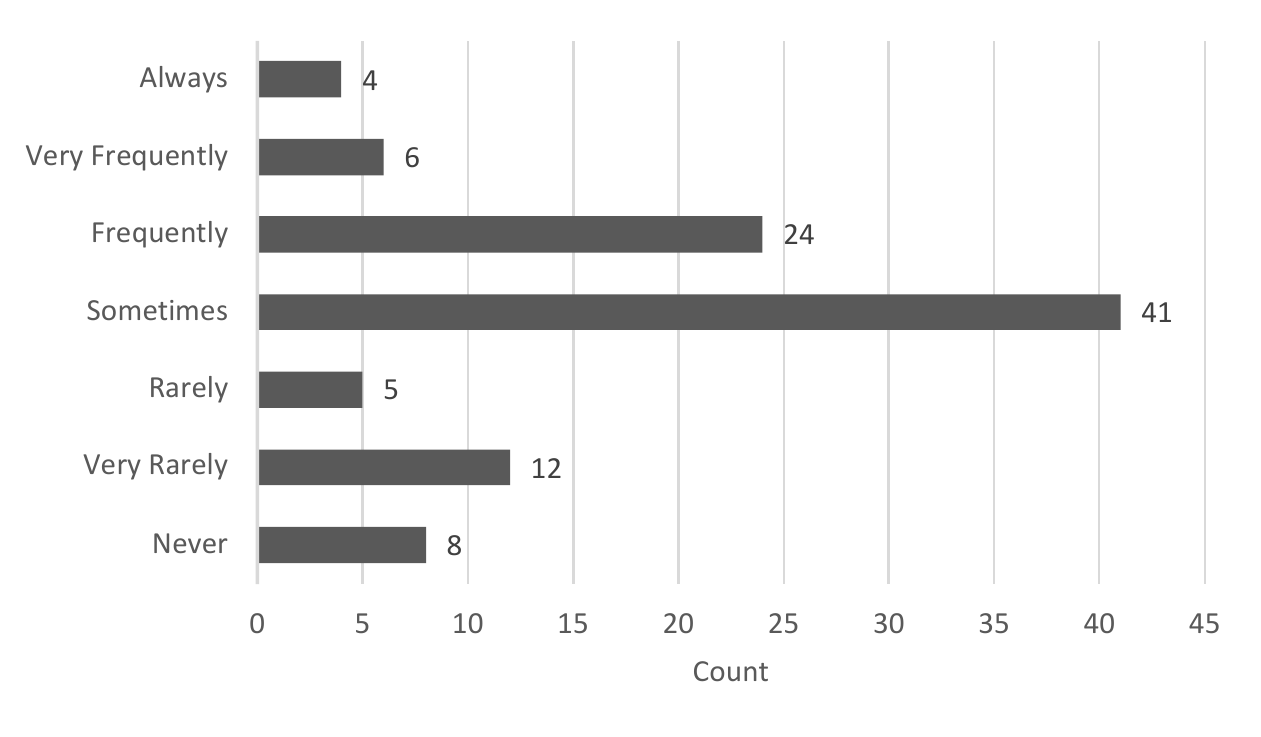}
\vspace{-10pt}
\caption{Did you encounter difficulties when engineering or maintaining self-adaptive systems? (Q4.1)}
\vspace{-10pt}
\label{fig:q4-1}
\end{figure}

Figure~\ref{fig:q4-1} summarises the findings. Forty-one of 100 participants report that they sometimes face difficulties with applying self-adaptation. Thirty encounter difficulties frequently or very frequently, while 17 rarely or very rarely have difficulties. Four participants reported to have always problems, while eight reported that they never face difficulties. 

\subsubsection{Please give one or two examples of the difficulties that you encountered when engineering or maintaining self-adaptive systems. (Q4.2)}

Seventy-four participants reported in total 140  difficulties, i.e., on average 1.9 difficulties per participant. Table~\ref{tab:codes_q4-2} summarises the findings.

\small
\small
\begin{table*}[hbt]
\caption{Analysis of comments -- Difficulties with engineering or maintaining self-adaptive systems (Q4.2)}
\label{tab:codes_q4-2}
\begin{tabular}{p{3.6cm}lp{8.8cm}}
\hline\noalign{\smallskip}
Categories and codes & \# & Example quotes\\
\noalign{\smallskip}\hline\noalign{\smallskip}

\textbf{Design issues} &  \textbf{43} &  \\

Reliable/optimal design & 26 & ``With high availability requiremets, the chance something fails somewhere sometime is close to a 100\%. The systems needs to be designed to still provide service despite erroneouse behavior or failing parts in the system.''; ``the main challenge is to design adaptation function with respect to computation context'' 
\\

Design complexity & 17 & ``Complexity in defining the adaptation rules. Conditions are not always obvious.''; ``Self-adaptiveness or resilience have to be taken into consideration at each stage of the 
... workflow. This is really a challenge as more often than not these are concepts that are completely obscure to the average programmer/devop mind.'' 
\\
\noalign{\smallskip}\hline\noalign{\smallskip}
\textbf{Lifecycle issues} & \textbf{42} &  \\

Tuning/debugging & 19 & ``Debugging the root cause of a scaling failure might be time-consuming: also, in some cases the problem might be outside of your control (e.g. temporary lack of EC2 Spot capacity in AWS)'' \\

Limitations tools/methods & 13 & ``The metrics available are not always fully transparent and built with auto-scaling in mind''; ``IAM permissions are hard to deal with when configuring these self-adaptive systems. Usually, the permission to scale or to notify is not properly configured.''\\

System/environment evolution & 10 & ``If the functionality is not designed in from the beginning then it is a huge amount of work to implement later.''; ``System architecture over lifetime (nee features to be added...)''
\\
\noalign{\smallskip}\hline\noalign{\smallskip}
\textbf{Runtime issues} & \textbf{30} &  \\

Runtime uncertainty & 17 & ``Many self-adaptive systems are based on unproven heuristics. Therefore, they usually do not work in many cases.''; ``It is hard to guess how much can the environment affect the system. ... 
It is hard to extend the parameters to cover whole production.'' \\

Data collection/evaluation & 7 & ``Gathering quantitative data samples to evaluate the performance is very complicated.''; ``sensors gives wrong reading values'' \\

Resources required & 3 & ``Sometimes it doesn't react fast enough. It also takes computation resources for this self-adaptive software, and the compute resources use increases with the number of incoming requests.'' \\

Delayed/missing runtime changes & 3 & ``Autoscaling is often too slow or triggered too late.''; ``Notifications are delayed or missed'' 
\\
\noalign{\smallskip}\hline\noalign{\smallskip}
\textbf{People and process issues} & \textbf{19} &  \\

Skills/experience & 14 & ``Every self-adapt system must be tuned up which is sometimes tricky and needs high skilled engineers.''; ``The Kubernetes/Openshift cloud and centralized log storage ... require experienced administration staff and vast knowledge of many networking concepts (... DNS, NAT).'' \\

Process and management & 9 & ``We are not yet very experienced ... the main challenges were to convince the central IT department this was the way to go, then to design the system, and obviously to master the technology itself.'' \\

Automation & 1 & ``often automation is not trusted enough by humans. humans want to stay in the loop.'' \\

\noalign{\smallskip}\hline
\end{tabular}
\end{table*}
\normalsize

\textbf{Analysis of comments:} We identified four categories of difficulties: \emph{design issues}, \emph{lifecycle issues}, \emph{runtime issues}, and \emph{people and process issues}. Most frequently reported difficulties, 43 in total, relate to the design of self-adaptation, in particular \emph{reliable/optimal design} (26 occurrences) and \emph{design complexity} (17 occurrences). Life cycle issues were reported 42 times, in particular \emph{tuning/debugging} (19 occurrences) and \emph{limitations of tools and methods} (13 occurrences). Difficulties with runtime aspects of self-adaptive systems was reported 30 times with \emph{runtime uncertainty} mentioned 17 times, and difficulties related to people and process occurred 25 times with \emph{skills and experience} occurring 14 times.  

\subsubsection{Did you face any risks when engineering self-adaptive systems? (Q4.3)}
\begin{figure}[h]
\centering
\includegraphics[width=0.7\columnwidth]{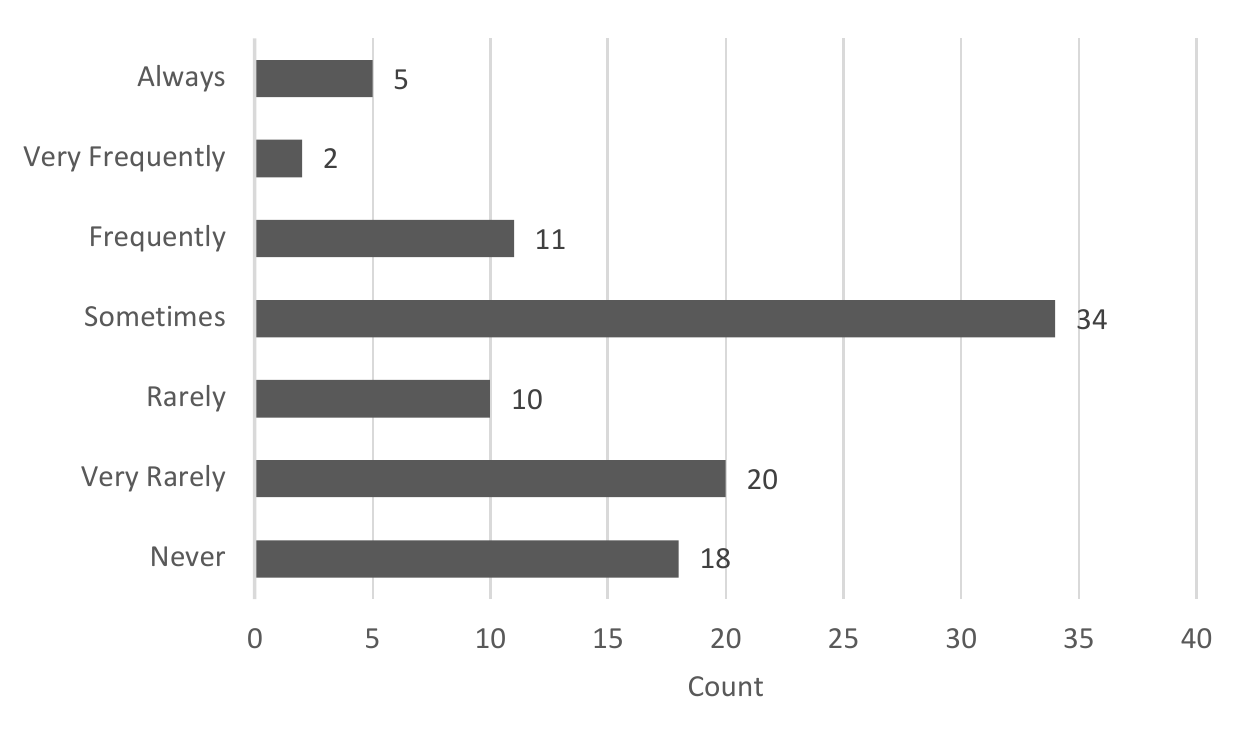}
\caption{Did you face any risks when engineering self-adaptive systems? (Q4.3) -- 100 answers}
\label{fig:q4-3}
\end{figure}

Figure~\ref{fig:q4-3} summarises the findings. 
Thirty-four of 100 participants report that they sometimes face risks when engineering self-adaptive systems. Eighteen report that they frequently to always encounter risks, while 48 rarely to never face risks. 

\subsubsection{Briefly describe one or two risks that you faced when engineering self-adaptive systems. (Q4.4)}

The participants provided a total of 60 responses containing 66 instances of risks faced when engineering self-adaptive systems.
On average, the participants reported 1.3 risks.

\textbf{Analysis of comments:} 
Tables~\ref{tab:codes_q4-4.I} and~\ref{tab:codes_q4-4.II} summarise the findings.
Out of the 60 valid answers, 11 were marked as unclear or not applicable.
We identified four categories of risks: \emph{faults}, \emph{difficulties with development/operation}, \emph{impact on qualities}, and \emph{impact on business}. 
Most frequently mentioned risks, 20 in total, relate to faults, in particular \emph{incorrect functionality} (7 occurrences), \emph{wrong results} and \emph{misconfiguration} (4 occurrences each), and \emph{network failure} (2 occurrences).
Difficulties with development/operation relate to \emph{difficulties to manage environment uncertainty} (6 occurrences), and \emph{difficulties to test} and \emph{build systems} (4 occurrences each). 
Participants mentioned also the risk of having several qualities impacted; \emph{performance degradation} with 5 occurrences the most frequent, followed by \emph{reduced availability} and \emph{safety and security threats} with 4 occurrences each.
Finally, negative impact on the business in terms of \emph{increased cost} (5 occurrences) and \emph{losing control and trust} (4 occurrences) are also reported as important risks when applying self-adaptation. 

\small
\begin{table*}[hbt]
\caption{Analysis of comments I -- Risks faced when engineering self-adaptive systems (Q4.4).}
\label{tab:codes_q4-4.I}
\begin{tabular}{p{3.6cm}lp{8.8cm}}
\hline\noalign{\smallskip}
Categories and codes & \# & Example quotes\\
\noalign{\smallskip}\hline\noalign{\smallskip}

\textbf{Faults} & \textbf{20} &  \\

Incorrect functionality & 7 & ``Automation can lead to unexpected values''; ``The process might be OOM killed if the self-adaptive system doesn't function correctly (i.e. bugs).'' \\
Wrong results & 4 & ``incorrect results''; ``Wrong decisions based on faulty models'' \\
Misconfiguration & 4 & ``Tuning autoscaling settings can be problematic resulting in unexpected results.''; ``Wrong threshold levels may lead to unwanted responses. '' \\
Network failure & 2 & ``Giving control to software that can change production environments can cause network failure.'' \\
Other & 3 & ``data loss''; ``[...] heuristics that work well on some applications, do not always perform the best for all applications.'' \\
\noalign{\smallskip}\hline\noalign{\smallskip}
\textbf{Difficulties with development/operation} & \textbf{16} &  \\

Difficult to manage environment uncertainty & 6 & ``We face a risk of underestimating environment variability.''; ``Legacy monitoring solutions don't cope well with environments that scale back.''; ``Risk may be encountered if the incoming event stream is completely unpredictable and have huge spike differences in data for a considerable period of timr'' \\
Difficult to test & 4 & ``if the executed actions that will be done by the self-adopting system are not tested before, it might introduce some risks''; ``It is also difficult to do reliable performance testing in non-production environments.'' \\
Difficult to build & 4 & ``implementing and designing self-adaptive systems may initially seem to take longer time – hence the risk of not being allowed to implement it as good as it can be done''; ``Costs of building own (self-hosted) environment [...]'' \\
Other & 2 & ``life updates (no downtime)''; ``There is always a lingering concern of quis custodiet ipsos custodes - or 'who watches the watchmen'.'' \\
 
\noalign{\smallskip}\hline
\end{tabular}
\end{table*}

\begin{table*}[hbt]
\caption{Analysis of comments II -- Risks faced when engineering self-adaptive systems (Q4.4).}
\label{tab:codes_q4-4.II}
\begin{tabular}{p{3.6cm}lp{8.8cm}}
\hline\noalign{\smallskip}
Categories and codes & \# & Example quotes\\
\noalign{\smallskip}\hline\noalign{\smallskip}

\textbf{Impact on qualities} & \textbf{16} &  \\

Performance degradation & 5 & ``[...] risk of degrading the performance instead of improving it, and degrading the user experience as a result.''; ``Performance impact on the running system when applying auto-scaling (e.g. scaling down)''; ``sometimes a sequence of perfectly acceptable self-adaptive automatic actions can lead to outages worse than the root cause'' \\
Reduced availability & 4 & ``If the system did not behave properly this could result in an outage [...]''; ``Availability of the system during the auto-scaling rules being applied''\\
Safety and security threats & 4 & ``If a system is self-adaptive, how can we secure that it is safe during production (some parts can be powered for self test during assembly and we need to know it is safe)? If we use machine learning on a self-adaptive system, how do we secure safety? 
''; ``There is a risk of misconfiguration that can lead to lost nodes and applications, security exposures etc. There are also security risks involved with the base building components, such as docker images from untrusted sources [...]'' \\
Extra resource consumption & 2 & ``Risk of all resources being eaten up by a self-adaptive process.''; ``[...] it may use up too many unnecessary hardware and software resources'' \\
Reliability issues & 1 & ``Reliability issues in case of non-converging oscillations or plain wrong output due to prolonged failures in the metrics collection pipelines or simply wrong algorithms'' \\
\noalign{\smallskip}\hline\noalign{\smallskip}
\textbf{Impact on business} & \textbf{14} &  \\

Increased cost & 5 & ``Regarding autoscaling, the main issue was to fail and so increasing the infra cost of the users due to bugs in the system.''; ``Lost control over system size. This also impacted the approx. total cost agreed with the customer.'' \\
Losing trust and control & 4 & ``Trust. Because flexible manufacturing systems have some kind of autonomous behavior with tasks that have been done manually, our clients are initially very sceptial and to not trust the systems initally''; ``risk of losing (manual) control of the system for the sake of automation'' \\
Harder to understand/fix & 3 & ``the whole system becomes more complex, hence fewer people understand all details of its behaviour.''; ``More difficult troubleshooting for a self-adapting, distributed system.'' \\
Not useful & 2 & ``The self-adaptive system might not perform better than the baseline when dealing with dynamic shapes, as the cost model might not be generic enough to predict the performance.'' \\
 
\noalign{\smallskip}\hline
\end{tabular}
\end{table*}
\normalsize

\subsubsection{How did you mitigate the risks that you faced? (Q4.5)}

The participants provided 51 responses containing 66 instances of risk mitigating techniques when engineering self-adaptive systems, i.e., 
on average 1.3 techniques per participant.

\textbf{Analysis of comments:} 
Table~\ref{tab:codes_q4-5} summarises the findings.
Out of the 100 valid answers, 13 were marked as unclear or not applicable.
The other participants mentioned a variety of risk mitigation mechanisms, which we grouped into three categories. 
\emph{Stakeholder-centred techniques} are the largest category with 25 occurrences, followed by \emph{offline techniques} and \emph{online techniques} with 18 and 9 occurrences each. 
Within stakeholder-centred techniques, \emph{rigorous design and development} (8 occurrences), \emph{code review} (4 occurrences), and \emph{human supervision} (4 occurrences) are the most popular risk mitigation techniques. 
\emph{Extensive testing} with 15 occurrences is the mostly mentioned offline technique, while \emph{runtime monitoring and analysis} with 6 occurrences is the mostly mentioned online technique to mitigate risks. 

\small
\begin{table*}[hbt]
\caption{Analysis of comments -- Techniques to mitigate risks when engineering self-adaptive systems (Q4.5).}
\label{tab:codes_q4-5}
\begin{tabular}{p{3.6cm}lp{8.8cm}}
\hline\noalign{\smallskip}
Categories and codes & \# & Example quotes\\
\noalign{\smallskip}\hline\noalign{\smallskip}

\textbf{Stakeholder-centered techniques} & \textbf{25} &  \\

Rigorous design and development & 8 & ``careful engineering so that there are open doors for manual intervention, when necessary, without lost of system availability nor hindering the automation mechanisms''; ``We try to have design sessions [...] and possibly enhance the design in the early phases of development''; ``Engineering analysis, testing, controlled deployment, ...'' \\
 
Code review & 4 & ``As always, planning, design reviews, code reviews, testing on several levels, monitoring the production.''; ``Each incident is taken into consideration and rules are always reviewed. '' \\
 
Human supervision & 4 & ``The responsibility was left to a human operator.''; ``Mainly by performing tests and human supervision (monitoring resource utilization)'' \\
  
Outsource & 3 & ``Outsource the cloud operation to a specialized provider (RedHat, AWS) where possible. In other cases, customers had to hire experienced administrators/go through extensive period of testing to gain the necessary experience.'' \\

Other (post mortem analysis, hiring experts, work in pairs, documentation) & 6 & ``When we hit a problem years after the fact, we perform a detailed post-mortem and try to think about other possible failures we may have missed.''; ``We hired (multiple) external consultancy firms to tap into their experience in deploying such a system.''; ``Work in pairs,
Document architectural decisions'' \\
\noalign{\smallskip}\hline\noalign{\smallskip}
\textbf{Offline techniques} & \textbf{18} &  \\

Extensive testing & 15 & ``test each action in isolation before it is provided to the system''; ``Automated and human testing. In addition for complex algorithms, we run parallel, correlated analysis.''; ``With automated and manual testing while injecting non-determinism to the test suite''; ``Extensive testing at the customers factory and fine tuning of the models.'' \\
 
Set operational boundaries & 2 & ``Defined max-amount of resources a system functionality/component is allowed to consume.''; ``Thresholds and some manual monitoring'' \\
 
Encryption  & 1 & ``State of the art encryption, encryption, and encryption.'' \\
\noalign{\smallskip}\hline\noalign{\smallskip}
\textbf{Online techniques} & \textbf{9} &  \\

Runtime monitoring and analysis & 6 & ``Alerts tracking high-level properties that can give us some assurance that the system is working fine.''; ``Monitor / review the automated actions.'' \\
 
Roll-out/roll-back strategies & 2 & ``Slow roll - only send the new system traffic in small increments (10\%, 20\%, ...) until production baselines are established for load, actual latency, etc.  This helped us determine what the MIN and MAX pod settings should be as well as VM heap sizes.''; ``Manual roll back to previous stable state of user profiles.'' \\
 
Run in non-business critical hours & 1 & ``We run our processes during the night, when there is less chance of interference with business critical (customer facing) systems.'' \\

\noalign{\smallskip}\hline
\end{tabular}
\end{table*}
\normalsize

\subsubsection{Have you faced or seen any problems
of self-adaptation for which you would
appreciate support from researchers (Q4.6)}
\begin{figure}[h]
\centering
\includegraphics[width=0.7\columnwidth]{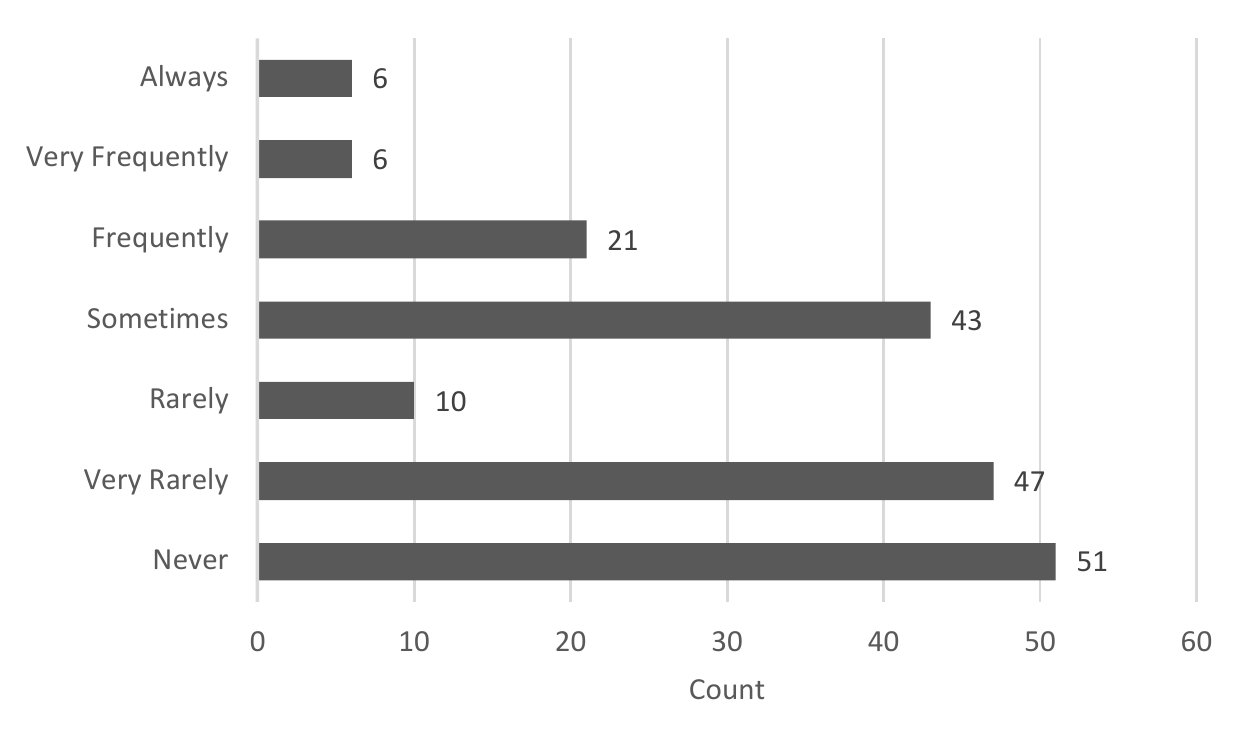}
\vspace{-10pt}
\caption{Have you faced or seen any problems of self-adaptation for which you would appreciate support from researchers? (Q4.6)}
\vspace{-10pt}
\label{fig:q4-6}
\end{figure}

Figure~\ref{fig:q4-6} summarises the findings for Q4.6. Thirty-three of 166 participants (17.9\%) frequently to always experience problems with self-adaptation for which they would appreciate support from researchers, while 43 participants (23.4\%) sometimes face such problems. On the other hand, 108 of the participants (58.7\%) never to rarely experience problems for which they would appreciate support from researchers. In summary, almost half of the participants believe that they would benefit from support of researchers to address some of the problems they face with engineering self-adaptive systems. 

\vspace{-10pt}




\subsubsection{For which problems of self-adaptation
would you appreciate support from researchers? Please briefly explain one or two such problems (Q4.7)}

Sixty-five participants described in total 113 problems for which they would appreciate support from researchers. Tables~\ref{tab:codes_q4-7.I} and~\ref{tab:codes_q4-7.II} summarise the findings.

\textbf{Analysis of comments:} 
We grouped the problems in four categories: \emph{engineering}, \emph{guarantees}, \emph{data}, and \emph{user interaction}. Forty-eight of the reported problems (42.5\% of the reported problems for which practitioners would appreciate support from researchers) relate to the engineering of self-adaptive systems. The main problems in this category relate to \emph{architecture and reuse} (16 occurrences) and the \emph{adoption} of self-adaptation (10 occurrences). Adoption refers to problems within a company with introducing self-adaptation, which can be related to technical aspects, expertise, or organisational aspect. Twenty-five of the reported problems (22.1\% of all reported problems) relate to guarantees, in particular providing \emph{trustworthiness} (20 occurrences) and dealing with \emph{unknowns} (five occurrences). Problems related to data were reported by 21 practitioners (18.6\%) and include \emph{data governance} and \emph{data access} (both eight occurrences), and \emph{machine learning} (five occurrences). The remaining 19 problems (16.8\%) relate to user interaction, namely \emph{automation} (nine times) and \emph{user experience} (seven times).   

\small
\begin{table*}[hbt]
\caption{Analysis of comments I -- Problems for which support of researchers would be appreciated (Q4.7)}
\label{tab:codes_q4-7.I}
\begin{tabular}{p{3.2cm}lp{9.2cm}}
\hline\noalign{\smallskip}
Categories and codes & \# & Example quotes\\
\noalign{\smallskip}\hline\noalign{\smallskip}

\textbf{Engineering} & \textbf{48} &  \\

Architecture \& reuse & 16 & ``Best Practices for implementation and architectural design guidelines''; ``I'd love to see a taxonomy of self-adaptive techniques. Perhaps a set of techniques could be added to Kazman's Architecture Tactics checklist?'' \\

Adoption & 10 & ``We lack interaction with development teams that are facing similar problems. We have a huge problem explaining this area to the management structure. ... they have basically no ability to lead due to lack of competence.''; ``new organisational structures and workflows that lead to the design of more self-adaptive and resilient platforms.'' \\

Platforms \& frameworks & 4 & ``to my knowledge there is no framework on what is 'safe' or not safe to be automatically executed by a self-adaptation system.''; ``To provide a platform for capturing the domain knowledge i.e. extensible ... to manage the managed systems what kind ... KPIs can be captured, and how they are related.''  \\

Tools & 4 & ``Outlier detection ... is well understood but existing commercial tools are usually pretty weak and custom code is required to optimize''; ``One of the main problems is to get tools that can profile the running systems under certain loads.'' \\

Testing \& debugging & 4 & ``Assurance of the behavior of highly dynamic systems is still the big hurdle. Test budgets and schedules do not grow with system complexity.''; ``a pre-production cloud test environment to try them first.''  \\

Advanced features & 10 & ``Coordinate multiple, potentially conflicting, objectives - in changing environment ... reacting too quickly [is] often sub-optimal''; ``research on network protocols, these should include some level of self-awareness and should automatically provide common network self-adaptation features.''; ``How a feedback loop can be designed in a way that you later can adapt to changes''  
\\
\noalign{\smallskip}\hline\noalign{\smallskip}
\textbf{Guarantees} & \textbf{25} &  \\

Trustworthiness & 20 & ``Formal verification of the algoritmic behaviour of the overall system (correctness)''; ``validate my algorithms''; ``Safety protocols for Machine learnign in self-adapting systems''; ``What are the mechanisms should be integrated into self-adapting system to identify malicious input?'' \\
Unknowns & 5 & ``We normally capture this using some form of process based models, but these struggle with thin[g]s like unknowns.''; ``not just anomaly detection, but actually responding appropriately to the anomalies (what is appropriate?).''  
\\

\noalign{\smallskip}\hline
\end{tabular}
\end{table*}

\begin{table*}[hbt]
\caption{Analysis of comments II -- -- Problems for which support of researchers would be appreciated (Q4.7).}
\label{tab:codes_q4-7.II}
\begin{tabular}{p{2.7cm}lp{9.7cm}}
\hline\noalign{\smallskip}
Categories and codes & \# & Example quotes\\
\noalign{\smallskip}\hline\noalign{\smallskip}

\textbf{Data} & \textbf{21} &  \\

Data governance & 8 & ``Data alignment and ... its integration''; ``getting data from application behaviour helps a lot in analyzing how application performance can be further improved.''; ``Adaptive AI systems to manage huge document contents'' \\

Data access & 8 & ``Support for data science as to extract correct cause relationships vs apparently correlations''; ``For example, how much data is shared across threads, how many objects are thread-local, how much performance is lost due to locality issues''  \\

Machine learning & 5 & ``if the data/metrics can be structured and labelled in some way (i.e. scored), then perhaps it should be possible to apply ML to help identify opportunities and figure out automatically how to respond.''; ``How to use machine learning to solve the self-adaptation problems and demonstrate its performance bound'' 
\\
\noalign{\smallskip}\hline\noalign{\smallskip}
\textbf{User interaction} & \textbf{19} &  \\

Automation & 9 & ``volume of data gets to large for people to process. People get to be the bottleneck for throughput''; ``Automatic synthesis of predictive and or reconfiguration models.''; ``Approaches whereby systems of reasonable scale can monitor and fix themselves as necessary without human intervention.''\\

User experience & 7 & ``most of the problems that we faced are related to help the customer to understand the benefits of self-adaptative systems.''; ``Autoscaling should become commodity products ... As users, the complexity should be abstracted away'' \\

User involvement & 3 & ``User response can also be used for adaption (E.G. if a user constantly overrides the managed systems settings there managing system should 'learn' from the user and adapt the control algorithm for that specific user)'' \\

\noalign{\smallskip}\hline
\end{tabular}
\end{table*}
\normalsize

\subsubsection{In your organisation or in industry in general, do you see application opportunities for self-adaptation that are currently not exploited? (Q4.8)}

Of the 184 participants, 101 (54.4\%) highlight new opportunities for applying self-adaptation, while 83 do not report any. The number of participants within these two groups is almost equally split among participants who have worked with concrete self-adaptive systems and those who have not (see Q0.1) (in particular, 58 participants that worked with self-adaptive systems report opportunities, while 42 do not).

\subsubsection{Please describe or give examples of the application opportunities for self-adaptation that are currently not exploited (Q4.9)}

Eighty-five participants described in total 147 unexploited opportunities for applying self-adaptation, i.e., an average of 1.7 opportunities per participant. 

\small
\begin{table*}[hbt]
\caption{Analysis of comments I -- Opportunities for self-adaptation that are not exploited yet (Q4.9)}
\label{tab:codes_q4-9.I}
\begin{tabular}{p{3cm}lp{9.4cm}}
\hline\noalign{\smallskip}
Categories and codes & \# & Example quotes\\
\noalign{\smallskip}\hline\noalign{\smallskip}

\textbf{System activities} & \textbf{72} &  \\

 Autonomous operation & 37 & ``E.g manufacturing production line with visual inspection operators who remove defects, ... the production line can further be adapted based on the defect rate/type''; ``Self adaption could have a lot of benefits in building automation systems, like smart heating and lighting systems that takes peoples habits into consideration.''; ``making the system adaptive to adjust and act instantly based on the data without waiting would be beneficial and efficient.'' \\
 
Data management \& machine learning & 26 & ``Methods to automatically handle changes in the machine learning models and to efficiently deploy them to the edge. There is still lots of manual fine tuning that delays a timely new release.''; ``The query optimizer of database (i.e. MySQL) could utilize self-adaptation technic.'' \\
 
 Autoscaling & 9 & ``The "managed service", which is a stateful service/ data store, is provisioned for the peak capacity, which means resources are idle most of the time. If we can build reliable and efficient system that can automatically scale stateful services based on the demand, we can reduce the cost.''; ``Our microservices do not dynamically scale'' \\
\noalign{\smallskip}\hline\noalign{\smallskip}
\textbf{System properties} & \textbf{47} &  \\

Quality improvement & 26 & ``Based on the alarm certain counter actions could be initiated in order to deal with the faulty behaviour and reach a stable system state.''; ``Congestion prognosis''; ``fault tolerance''; ``Power consumption''; ``resource optimization''; ``There are many opportunities to split up [current monolithic systems] and then make them scalable such that outages are more contained. E.g. screens on trains.'' \\

Security improvement & 10 & ``Security of e.g., mobile devices that adapts based on locally identified threats as well as knowledge of risks in the environment.''; ``Automating changes in Security levels based on threat levels''; ``Detecting in-vehicle threats, detecting a system being compromised''; ``react to attack patterns'' \\

Cost effectiveness & 8 & ``IT cost reduction (e.g. software asset mgmt)''; ``The question really is: How do you do these things on the cheap (with non Silicon Valley billion dollar funding) and in contexts where mistakes might be extremely critical?'' \\

\noalign{\smallskip}\hline
\end{tabular}
\end{table*}

\begin{table*}[hbt]
\caption{Analysis of comments II -- Opportunities for self-adaptation that are not exploited yet (Q4.9).}
\label{tab:codes_q4-9.II}
\begin{tabular}{p{3.1cm}lp{9.4cm}}
\hline\noalign{\smallskip}
Categories and codes & \# & Example quotes\\
\noalign{\smallskip}\hline\noalign{\smallskip}

\textbf{Engineering activities} & \textbf{21} &  \\

Maintenance \& reuse & 15 & ``self-adapting CI/CD infrastructure based on demand''; ``Preventive maintenance''; ``Carriers are eager to get rid of human factors to improve operation and maintenance capabilities and network quality. Therefore the ICT field pays much attention to self-adaption systems.''; ``Software provisioning and automatic updates'' \\

Patterns \& libraries & 6 & ``Developing a comprehensive library of algorithms on top of the industrial monitoring systems which can be applied to analysis portion of the chain in order to drive correct self-adaptation actions would benefit the self-adaptation adoption.''; ``Cross-cloud self-adaptation''; ``patterns to provide solutions to common problems'' \\
\noalign{\smallskip}\hline\noalign{\smallskip}
\textbf{Human involvement} & \textbf{7} &  \\

Personalization & 4 & `` it would be interesting to adapt the player experience itself based on the player, mostly to better challenge them''; ``Healtcare decision making systems witch are changing outcomes and advices basd on patient status.'' \\

Human-machine interaction & 3 & ``I consider that the biggest opportunities are found within the Human Machine Interaction or Building Machine Interaction. There will be a future in which talking to a device that can modify the environment (e.g. a robot but not a phone) will be as natural as talking to a person, or seeing a machine interacting with another machine (e.g. robot taking the elevator)'' \\

\noalign{\smallskip}\hline
\end{tabular}
\end{table*}
\normalsize

\textbf{Analysis of comments:} 
Tables~\ref{tab:codes_q4-9.I} and~\ref{tab:codes_q4-9.II} summarise the findings.
We grouped the opportunities in four categories: \emph{system activity}, \emph{system property}, \emph{engineering activity}, and \emph{human involvement}. Seventy-two of the reported opportunities (i.e., 49\% of all) are related to system activity. The opportunities in this category relate to the \emph{autonomous operation} behaviour of self-adaptive systems (37 occurrences), \emph{data management and machine learning} (26 occurrences), and \emph{auto-scaling} (nine occurrences). Forty-seven opportunities (32\%) are related to system properties. In this category, the opportunities are related to \emph{quality improvement} (26 occurrences), 
\emph{security improvement} (10 occurrences), and \emph{cost effectiveness} (eight occurrences). Twenty-one of the reported opportunities (14.3\%) relate to engineering activities, in particular \emph{maintenance and reuse} (15 occurrences), and \emph{patterns and libraries} (six occurrences). Finally, seven opportunities (4.8\%) relate to human involvement, in particular \emph{personalisation} (four occurrences) and \emph{human-machine interaction} (three occurrences).  

\begin{framed}
\noindent \textbf{Key insight(s) from RQ4:} 
\begin{enumerate}
\item A majority of participants face difficulties when engineering or maintaining self-adaptive systems, mainly with reliable/optimal
design, design complexity, and tuning/debugging. 
\item About half of the participants encounter risks when using self-adaptation. The main risks relate to incorrect functionality and difficulty to manage environment uncertainty, as well as degraded performance and increased cost.
\item About half of the practitioners report that they would appreciate support from researchers to deal with problems they face, in particular problems related to the engineering of self-adaptive systems, guarantees, and management of data. 
\item About half of the participants see future opportunities for applying self-adaptation, in particular in relation to autonomous operation, data management and machine learning.  
\end{enumerate}
\end{framed}

\subsection{Confidence}
\label{subsec:rq5-1}

\begin{figure}[h]
\centering
\includegraphics[width=0.85\columnwidth]{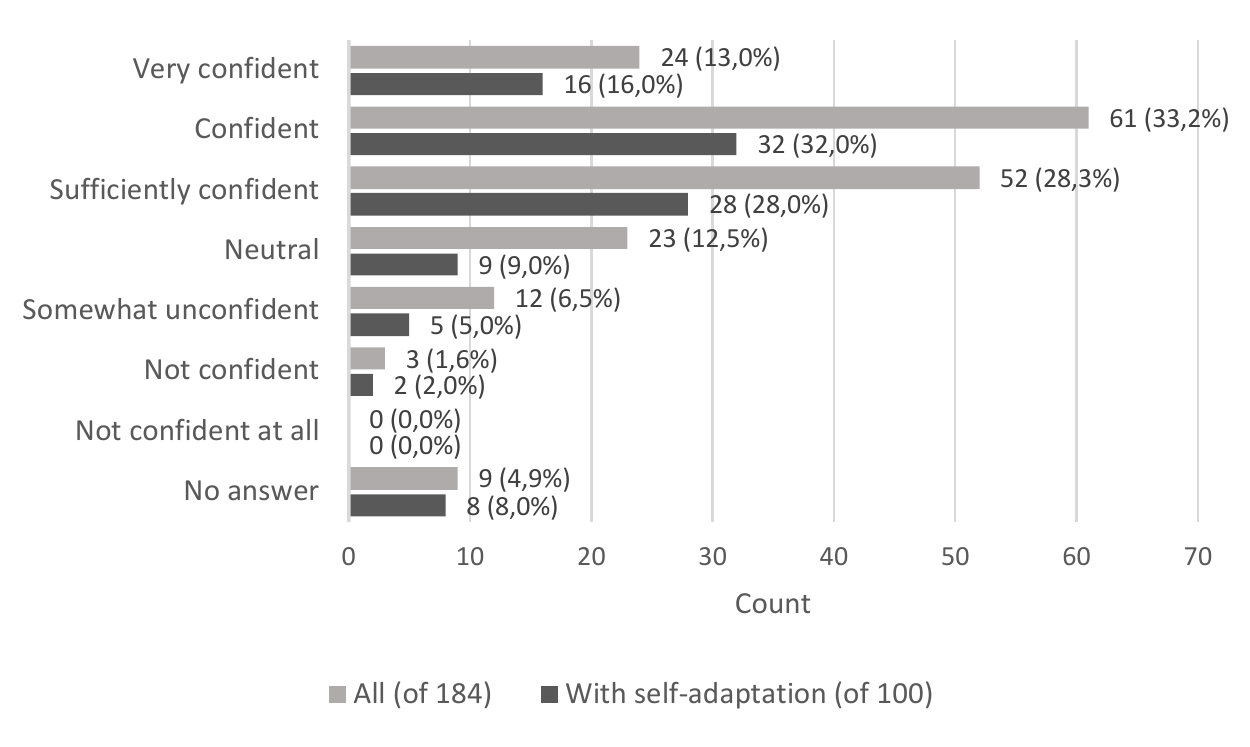}
\vspace{-10pt}
\caption{Confidence in answers.}
\vspace{-10pt}
\label{fig:q5-1}
\end{figure}

Figure\,\ref{fig:q5-1} shows the answers about how confident participants were in general about the answers they gave when answering the survey questions. The results show that almost all participants have confidence in the answers they provided to the survey questions. The numbers for all participants and those that have worked with self-adaptation are similar. 

\section{Discussion}
\label{sec:cross_analysis}

We start the discussion with highlighting a number of observations that we derived from the data analysis. Then we perform a number of additional analysis based on cross analyses of selected data of the answers of different questions. With this cross analyses we aim to gain further insights into three topics of interest: benefits of applying self-adaptation in practice, difficulties and risks with engineering self-adaptation in practice, and research support to address problems in practice. 

\subsection{Observations}

The problems addressed by industry are in general similar to those studied by academics. Yet, one particular difference is the lack of emphasis of practitioners on the use of self-adaptation to mitigate uncertainties, which has been a key focus in research\,\cite{garlan2010seu,Relax2010Whittle,esfahani2013usa,10.1145-3487921}. A possible explanation is that practitioners avoid the term uncertainty that may be perceived as ``doubt,'' ``not clearly defined,'' or ``not under control.'' Instead, they refer to uncertainty indirectly by using a different vocabulary, such as ``conditions are not always obvious'' and ``available metrics are not always fully transparent.''

While practitioners apply self-adaptation to deal with a variety of problems, changes in business goals are less frequently solved by using self-adaptation. One possible explanation may be that business goals are usually about higher-level requirements, while the focus of self-adaptation is often targeting ``lower level''  technical problems. In addition, there is also the challenge of the mapping between business goal and technical/system metrics, which touches the line or work on dynamic software product lines\,\cite{4488260}. Yet, another explanation may be that self-adaptation has not yet been fully utilised in industry to deal with bigger system changes. We hypothesise that the latter is the case, but further study is needed to obtain deeper insight. 

The four classic management tasks of self-adaptation studied by researchers (self-healing, self-optimising, self-protecting, and self-configuring) are also relevant to practitioners. Yet, differently from academics, practitioners also emphasise the importance of improving user satisfaction, reducing costs, and mitigating risks.  

Practitioners make extensive use of tools and infrastructures to realise the different functions of self-adaptation. This points to the need for more emphasis on tools and supporting infrastructure in research. Related to that is the need for reusing solutions, for instance in the form of references architectures and patterns. While some research efforts have been taken in these directions, these issues deserve more attention. An interesting step in this direction is the development of industry relevant artifacts as outlined in\,\cite{3561846.3561852}. 

Self-adaption in software-intensive systems is often not completely automated. Humans remain involved in adaptation, either to provide parts of functions or just to supervise the system. On the one hand, for some companies this is the first step towards further automation; on the other hand, practitioners often express the need for involving humans to ensure trust by overseeing the system and take action when something unexpected happens. As such, we expect the role of humans in self-adaptation to remain important also for future industry relevant research in self-adaptation. 

It is remarkable that more than 50\% of the participants report that they face at least sometimes risks with applying self-adaptation. At the same time, about half of the practitioners express that they would appreciate support from researchers
to deal with the problems they face. This suggests that the engineering of efficient and trustworthy self-adaptive systems is a challenge in practice and that practitioners believe that support from research could benefit them to deal with these challenges. This opens opportunities for joint efforts between industry and academics.

\subsection{Benefits of Applying Self-Adaptation in Practice}


When we crosscheck adaptation problems (Q1.1) versus kind of systems (Q0.2), we observe that most adaptation problems are applied to all kind of systems, while each adaptation problem is applied in one or two champion kind of systems. The three most frequently addressed adaptation problems are applied by all kind of systems. Specifically, the problem ``to optimise system performance'' is applied to all kinds of systems except transportation where ``to detect and resolve errors'' is the main adaptation problem (six occurrences), finances where ``to deal with changes in the environment'' is the main problem (five occurrences), and manufacturing where ``to automate tasks'' is the main problem (seven occurrences). Table\,\ref{tab:q1.1-q0.2} summarises the top occurrences, i.e., types of adaptation problems solved (top occurrences) versus the kind of system for which that adaptation problem is applied (top kind of systems).

\small
\begin{table}[hbt]
\centering
\caption{Cross analysis adaptation problem solved (Q1.1) versus kind of systems (Q0.2)}
\label{tab:q1.1-q0.2}
\begin{tabular}{lc}
\hline\noalign{\smallskip}
\textbf{Adaptation problem (top occurrences/total)} & \textbf{Top kind of system} \\
\noalign{\smallskip}\hline \noalign{\smallskip\smallskip}
To optimise system performance (12/78) & Embedded/cyber-physical/IoT \\
To automate tasks (10/61) & Cloud \\
To deal with changes in the environment (9/60) & Embedded/cyber-physical/IoT \\
To detect and resolve errors (8/46) & Web/mobile\\
To configure/reconfigure a system (8/51) & Web/mobile and Cloud\\
To detect and protect a system against threats (6/46) & Web/mobile\\ 
To deal with changes in business goals (5/15) & ICT communication and networks \\ 
\noalign{\smallskip}\hline
\end{tabular}
\end{table}
\normalsize

We now look at the problems for which self-adaptation is applied (Q1.1) versus the benefits of using self-adaptation (Q1.2). 
Table\,\ref{tab:contingency_q1-1_q1-2} shows the contingency matrix. The results show that ``improving user satisfaction'' and ``reducing  costs'' are by far the most frequently perceived benefits across all types of problems solved with self-adaptation. In particular, these two benefits are mentioned approximately 70\% (+/- 4\%) on average across all problems, while ``mitigating risks'' and ``penning up new application opportunities'' are respectively mentioned 53\% (+/- 11\%) and 28\% (+/- 5\%) on average across all problems solved with self-adaptation.

\small
\begin{table}[hbt]
\centering
\caption{Contingency matrix adaptation problem (Q1.1) versus benefits (Q1.2)}
\label{tab:contingency_q1-1_q1-2}
\begin{tabular}{lcccc}
\hline\noalign{\smallskip}
Problem/Benefit & Improve user satisfaction & Reduce costs & Mitigate risks & New opportunities\\
\noalign{\smallskip}\hline\noalign{\smallskip}
Automate tasks         & 42 & 45 & 34 & 15 \\
Environment changes    & 43 & 44 & 28 & 17 \\
Optimise performance   & 12 & 10 &  6 &  5 \\ 
Changes business goals & 55 & 55 & 35 & 17 \\ 
Handle errors          & 34 & 32 & 27 & 12 \\ 
Protect system         & 22 & 23 & 24 & 11 \\ 
(Re-)configure system  & 35 & 36 & 26 & 16 \\ 
\noalign{\smallskip}\hline
\end{tabular}
\end{table}
\normalsize

Finally, we look at the potential benefits of reuse using the data of the kind of software systems built by organisations (Q0.2) versus reuse when applying self-adaptation (Q3.5-3.7). 
The top domains where solutions are frequently reused are data management with 11 occurrences and embedded/cyber-physical/IoT systems with seven occurrences. Manufacturing is the top domain where practitioners very frequently reuse solutions with seven occurrences. The most frequent type of reused artifact is module with 11 occurrences, with embedded/cyber-physical/IoT as the top domain with four occurrences used for monitoring/analytics/control. Overall, there is no specific artifact that is more reused than other, and no domain that clearly reused more or less artifacts. Only five  participants mention the reuse of patterns when engineering solutions for self-adaptation. 

\begin{framed}
\noindent \textbf{Summary for Benefits of Applying Self-adaptation in Practice.} Optimising performance and dealing with changes in the environment are the main reported problems solved using self-adaptation in the domain of embedded/cyber-physical/IoT. Not surprisingly, self-adaptation in the cloud is primarily used to automate tasks and reconfigure the system. 
Reuse of self-adaptation solutions is mostly applied in the domains of manufacturing, data management, and embedded/cyber-physical/IoT systems. The main artifact of reuse is system module. 
\end{framed}

\subsection{Difficulties and Risks of Applying Self-Adaptation in Practice}


Large and small/medium organisations (Q0.3) are equally concerned about difficulties with design (Q4.1-4.2). Both types of companies are also concerned about tool support, but in different ways:  difficulties with debugging is more important for large organisations, while limitations of tools and methods more important for small/medium organisations.

When comparing large companies (>100) and small/medium companies (<100) (Q0.3), we observe no major difference in the reported frequency of encountered risks (Q4.3-4.4). The only relevant difference is that larger companies mention faults twice as much as small/medium ones; 14 occurrences for 30 large companies versus six for 70 small/medium companies. 

To crosscheck size of companies (Q0.3) versus mechanisms used to realise self-adaptation (Q3.1-3.3), we performed a dedicated coding distinguishing mechanisms that rely on tools/infrastructure versus custom mechanisms. 
%
%
The data summary shown in Table~\ref{tab:contingency_q0-3_q3-1_q3-3} indicates that smaller/medium companies (<100) rely on tools and infrastructure to provide support for self-adaptation mechanisms, while in large companies (>100) custom solutions are more prevalent. 
Zooming into the data of mechanisms for the different stages of self-adaptation shows that almost all companies that apply self-adaptation have mechanisms in place for monitoring, but not necessarily for analysis and change, regardless of company size, but the differences are small. This suggests a progression from monitoring to analysis to change.  

\small
\begin{table}[hbt]
\centering
\caption{Contingency matrix size of companies (Q0.3) versus self-adaptation mechanisms (Q3.1-3.3)}
\label{tab:contingency_q0-3_q3-1_q3-3}
\begin{tabular}{lcc}
\hline\noalign{\smallskip}
Size company & Relying on tools/infrastructure & Custom mechanisms\\
\noalign{\smallskip}\hline\noalign{\smallskip}
1-10     & 5 (56\%) & 4 (44\%) \\
11-20    & 3 (27\%) & 8 (73\%) \\
21-50    & 5 (36\%) & 9 (64\%) \\ 
51-100   & 4 (40\%) & 6 (60\%) \\ 
$> 100$  & 7 (13\%) & 47 (87\%) \\ 
\noalign{\smallskip}\hline
\end{tabular}
\end{table}
\normalsize

Cross analysis of subject of adaptation (Q2.1) versus difficulties and risks (Q4.1-4.2) shows that the reported difficulties and risks are similarly distributed across subjects of adaptation. Most frequently reported difficulties are design issues and
people and process issues at system level (both 11 instances). Most frequently reported risks are difficulties development/operation and impact on business also at system level (six and five occurrences, respectively).  
\begin{framed}
\noindent \textbf{Summary for Difficulties and Risks with Engineering Self-adaptation.} The main difficulties concern the design of self-adaptation and people and processes at system level, while the main risks relate to development/operation and impact on business, also at system level. Large companies face higher risks related to faults when applying self-adaptation. Difficulties with design is important for all, yet, debugging is more important for large companies, while small/medium companies are more concerned about limitations of tools and methods. 
\end{framed}

\subsection{Research Support to Address Problems in Practice}


Table\,\ref{tab:q2.1-q4.6} shows the main results of the cross analysis of the data of the concrete self-adaptive systems built by the participants (Q2.1) and the problems for which practitioners would appreciate, sometimes to always, support from researchers (Q4.6).   

\small
\begin{table}[hbt]
\centering
\caption{Cross analysis concrete self-adaptive systems (Q2.1) built vs support from researchers (Q4.6)}
\label{tab:q2.1-q4.6}
\begin{tabular}{lc|lc|lc}
\hline\noalign{\smallskip}
\textit{Subject adaptation} &  Support & \textit{Type adaptation} & Support & \textit{Trigger adaptation} & Support \\
\hline\noalign{\smallskip}
System & 12 (26.7\%) & Auto-tuning & 16 (36.4\%) & System properties & 11 (31.4\%) \\
Module & 9 (20.0\%) & Auto-scaling & 13 (29.5\%) & Environment properties & 8 (22.9\%) \\
Application layer & 9 (17.8\%) & Monitor/analysis & 9 (20.5\%) & System load & 6 (17.1\%) \\
\noalign{\smallskip}\hline
\end{tabular}
\end{table}
\normalsize

The analysis shows that system, module and application layer make a total of 64.4\% of the problems for which practitioners would appreciate support from researchers. In terms of type of adaptation, 84.6\% of the problems for which practitioners would appreciate support from researchers concern auto-tuning, auto-scaling, and monitoring and analysis. Finally, 74.1\% of the problems for which support would be appreciated concern adaptation triggered by system properties, environment properties, and system load.   

When crosschecking the kind of software systems built by the practitioners (Q0.2) versus the problems for which they would appreciate at least sometimes support from researchers (Q4.6), we found that except for one kind of system, support from researchers would be appreciated across all kinds of systems built by the practitioners. For e-commerce none of the seven participants expressed interest in regular support from researchers (four of them would very rarely appreciate support). On the other hand, eight out of 11 (72.2\%) participants that work in the domain of ICT communication and networks would regularly appreciate support to address their problems. The numbers for the other domains range from 22.9\% to 56.3\%. 

\begin{framed}
\noindent \textbf{Summary for Research Support to Address Problems in Practice.} A majority of practitioners would appreciate support from researchers. These problems concern self-adaptation applied at system level, a module of the system, or the application layer. The main problems relate to auto-tuning, auto-scaling, and monitoring and analysis. Triggers of adaptation concern dealing with system and environment properties, and system load. The problems crosscut different kinds of systems, but particularly ICT communication and networks.  
\end{framed}

\section{Threats to Validity }\label{sec:ThreatsToValidity}

We discuss validity threats of our study using the guidelines described in~\cite{Wohlin2012}. We look at construct validity that refers to the extent to which we obtained the right measure and whether we defined the right scope for the study goal, external validity that refers to the extent to which the findings can be generalized, and reliability that refers to the extent to which we can ensure that our results are the same if our study is done again. 

\subsection{Construct Validity} 

The survey starts from the assumption that practitioners are sufficiently familiar with the basic concepts of self-adaptation. We used the term self-adaptation to formulate questions about systems (or parts) that are equipped with a feedback loop. Hence, most questions required basic knowledge of the concept of self-adaptation. Analysis of the results makes it clear that practitioners have a basic understanding these concepts. Yet, we used several measures to reduce possible misinterpretations. We introduced the notion of self-adaptation at the start of the survey using a standard model with a feedback loop that we illustrated with typical examples. We elicited feedback for several participants on this description and the questions during a pilot. This feedback enabled us to enhance the description and clarify some of the questions. In addition, we selected participants with sufficient experience from a variety of domains. The confidence in the answers (Q5.1) confirms that the participants believed that their answers were trustworthy.   

\subsection{External Validity}

A potential threat to validity may be the generalisation of the study results.  Core to this threat is the selection of the sample of the target population. If this population may not have been representative, the study results may be imprecise and hence not generalisable. Since we used a non-probabilistic sampling method, there is a potential risk that the sample used to conduct the survey is biased and not representative of the target population. To mitigate the validity threat we mainly reached out to practitioners from our networks with industry. To ensure that participants have the required experience, we included questions asking about personal experience with engineering self-adaptive systems in practice. The results of the demographics of our sample show that the participants were active practitioners with sufficient expertise in various roles across companies of different sizes.  In addition, we worked in total with eight teams from different areas that contacted practitioners from all over the world. This ensured a well-balanced population on a global scale. Because several of the researchers involved in this study are active in the field of engineering self-adaptive systems, the practitioners invited from our networks may have been biased and inclined to apply self-adaptation more often. To anticipate this threat, we did not particularly focus on practitioners that we have worked within projects, but rather invited practitioners in various software engineering roles that are active across a wide range of domains.  

\subsection{Reliability}
 
Data analysis, in particular qualitative analysis (coding of answers with free text), are creative tasks that are to some extent subjective. Performing these tasks may be influenced by the experience (and even opinions) of the coders~\cite{Fernandez2016}. To mitigate a potential interpretation bias, we followed a thorough coding scheme.  The coding tasks were distributed among teams of two authors (one team of three). The authors of each team performed coding of the data independently and discussed where needed until an agreement was reached. All coding tasks were then distributed again among two authors. These authors repeated the coding independently from the initial coding. The results were then compared with the initial coding by these two authors. Any discrepancies were discussed among the two authors until consensus was reached. The coding was finally crosschecked with the authors that did the original coding to reach consensus. Finally, all material of the survey, including the raw data and the coding are publicly available.\footnote{https://people.cs.kuleuven.be/danny.weyns/surveys/sas-in-industry/}

\section{Conclusions}
\label{sec:conclusions}
In this paper, we studied the application of self-adaptation in industry. To that end, we conducted a questionnaire-based survey with practitioners from all over the world. We received valid responses from 184 participants, 100 of them with experience in engineering self-adaptive systems. 

\begin{figure}[h]
\centering
\includegraphics[width=1.0\columnwidth]{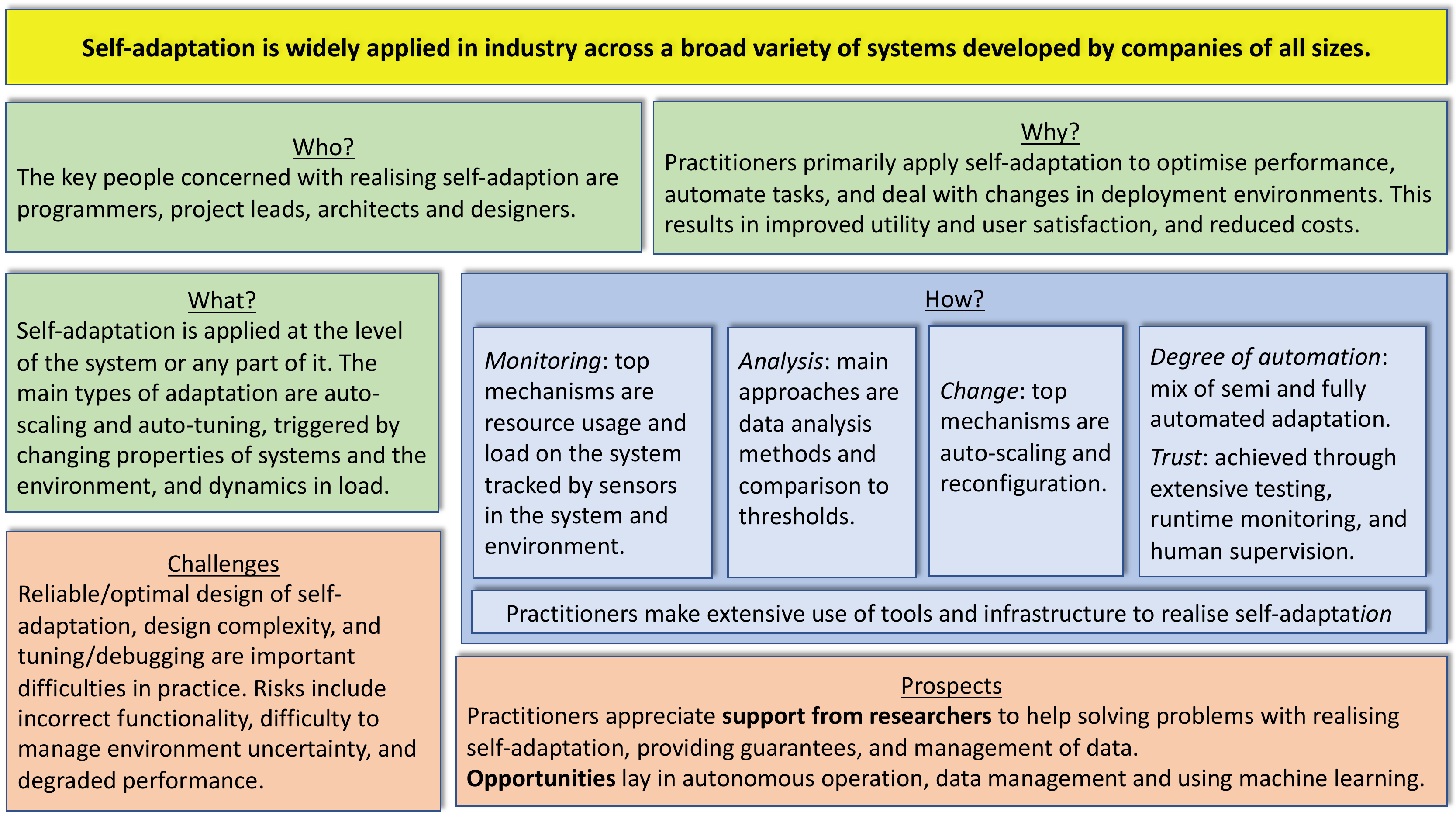}
\caption{Summary of the main findings of the survey}
\label{fig:summary}
\end{figure}

By analysing the data, we contributed an empirically grounded overview of state-of-the-practice in the application of self-adaptation. A selection of key observations includes: i) self-adaptation is extensively applied in industry across a wide variety of domains, ii) the dominating types of adaptations applied in industry are auto-scaling, auto-tuning, and monitoring/analysis, iii) practitioners rely extensively on tools and infrastructure to realise the different functions of self-adaptation, iv) human supervision is important to ensure trust in industrial self-adaptive systems, v) about half of the participants encounter risks with applying self-adaptation, vi) on the other hand, about half of the practitioners would appreciate support from researchers to deal with problems they face. Figure\,\ref{fig:summary} summarises the main findings. 

The results offer insights for researchers that enable them to  compare the focus their of their current research with industrial needs. A selection of related key insights includes: i) different from academics that study adaptation for mitigating uncertainty of classic maintenance tasks (self-*), practitioners also emphasise the importance of improving user satisfaction, reducing costs, and mitigating risks, ii) practitioners (in particular those of small and medium sized companies) rely on tools and infrastructure to realise self-adaptation, iii) ensuring trust in industrial self-adaptive systems is mainly achieved through extensive testing, runtime monitoring and alerting, and human supervision, iv) risks with self-adaptation in practice relate mainly to incorrect functionality, difficulty to manage environment uncertainty, 
degraded performance and increased cost.

The results also offer insights for practitioners to assess the level of their current practice in applying self-adaptation. A selection of related key insights includes: 
i) practitioners broadly confirm that the use of self-adaptation improves robustness and performance while reducing costs and required resources, and improves user experience while reducing the burden of engineers, ii) a wide range of mechanisms are used to enact self-adaptation in industrial systems, iii) tools and infrastructure, such as auto-scaling and container-orchestration platforms are available and commonly used to support the realisation of self-adaptation in practice, iv) important challenges when engineering self-adaptation in practice are reliable/optimal design, design complexity, and tuning/debugging, v) there is a relevant match between industrial practice in realising self-adaptation and the body of work performed by the research community of self-adaption.    

The survey results provide prospects for applying self-adaptation in practice and opportunities for industry-research collaborations in this area. The prospects include: i) realising full autonomous operation, ii) exploiting machine learning, iii) improving quality and security, and iv) applying self-adaptation for maintenance. Key opportunities for industry-research collaborations are in: i) consolidating best practices (architecture, patterns, and reuse), ii) modelling paths for the adoption of self-adaptation in industry, iii) supporting advanced features to realise self-adaptation such as dealing with the evolution of self-adaptive systems, iv) rigorous methods for ensuring trustworthiness of self-adaptive systems, v) governance of data, and vi) moving the human in the loop (performing adaptation functions) to the human on the loop (overseeing the system to ensure trust).  

We hope that the results of this survey will propel industry-relevant research in the field of self-adaptive systems and enhance the application of self-adaptation in practice, paving the way for self-adaptation to reach full maturity as a discipline. 

\section*{Acknowledgement}

We would like to thank the participants of our study and the reviewers of the survey protocol.

\bibliographystyle{ACM-Reference-Format.bst}
\bibliography{references}


\end{document}